\documentclass[a4paper,openany, 12pt]{book}

\usepackage{amsfonts,amsmath,amssymb,amsthm}
\usepackage{bm}
\usepackage{enumitem}
\usepackage[text={15.5cm,23cm},centering]{geometry}
\usepackage{graphicx}
\usepackage[colorlinks=true,linkcolor=linkcolour,citecolor=red,linktocpage=true,breaklinks=true]{hyperref}
\usepackage{mathptmx}
\usepackage[framemethod=default]{mdframed}
\usepackage{tcolorbox}
\usepackage{tikz} 
\usepackage{url}
\usepackage{varwidth}

\RequirePackage[framemethod=default]{mdframed} 

\newcounter{dummy-example}
\newcounter{dummy-exercise}  

\numberwithin{dummy-example}{chapter}
\numberwithin{dummy-exercise}{chapter}

\newtheoremstyle{maincolornumbox}
{0pt}
{0pt}
{\normalfont}
{}
{\small
	\sffamily\color{maincolor}}
{\;}
{0.25em}
{\sffamily\bfseries\color{maincolor}\thmname{#1} \thmnumber{#2}\thmnote{\sffamily\mdseries\color{black}{ ---}\nobreakspace #3.}\newline\noindent} 

\theoremstyle{maincolornumbox}
\newtheorem*{proofT*}{Proof}
\newtheorem{exampleT}[dummy-example]{Example}
\newtheorem{exerciseT}[dummy-exercise]{Exercise}

\newmdenv[skipabove=7pt,
skipbelow=7pt,
backgroundcolor=lightgrey,
linecolor=maincolor,
innerleftmargin=5pt,
innerrightmargin=5pt,
innertopmargin=5pt,
leftmargin=0cm,
rightmargin=0cm,
innerbottommargin=5pt,
linewidth=1]{tBox}

\newmdenv[skipabove=7pt,
skipbelow=7pt,
rightline=false,
leftline=true,
topline=false,
bottomline=false,
backgroundcolor=lightgrey,
linecolor=maincolor,
innerleftmargin=5pt,
innerrightmargin=5pt,
innertopmargin=5pt,
innerbottommargin=5pt,
leftmargin=0cm,
rightmargin=0cm,
linewidth=3pt]{eBox}	

\newmdenv[skipabove=7pt,
skipbelow=7pt,
rightline=false,
leftline=true,
topline=false,
bottomline=false,
linecolor=maincolor,
backgroundcolor=maincolor!10,
innerleftmargin=5pt,
innerrightmargin=5pt,
innertopmargin=5pt,
leftmargin=0cm,
rightmargin=0cm,
linewidth=3pt,
innerbottommargin=5pt]{dBox}

\newmdenv[skipabove=0pt,
skipbelow=5pt,
rightline=false,
leftline=false,
topline=false,
bottomline=false,
backgroundcolor=lightgrey,
linecolor=maincolor,
innerleftmargin=5pt,
innerrightmargin=5pt,
innertopmargin=-8pt,
innerbottommargin=5pt,
leftmargin=0cm,
rightmargin=0cm]{eqBox}		

\newenvironment{example}{\begin{dBox}\begin{exampleT}}{\end{exampleT}\end{dBox}}

\newcounter{ex-num}
\newenvironment{exercises}{
	\begin{eBox}
		\textcolor{maincolor}{\textsf{\textbf{Exercises}}} 
		\begin{enumerate}[label={\thechapter.\arabic*},topsep=2pt]
			\setcounter{enumi}{\value{ex-num}}
		}{
			\setcounter{ex-num}{\value{enumi}}
		\end{enumerate}
	\end{eBox}
}	

\newcommand{\ba}{\begin{eqnarray}}
\newcommand{\ea}{\end{eqnarray}}
\newcommand{\be}{\begin{equation}}
\newcommand{\ee}{\end{equation}}
\newcommand{\beq}{\begin{equation}}
\newcommand{\eeq}{\end{equation}}

\newcommand{\bba}{\begin{eqBox}\begin{eqnarray}}
\newcommand{\eba}{\end{eqnarray}\end{eqBox}}

\newcommand{\tr}{\operatorname{tr}}

\newcommand{\identity}{\mathbb{I}}

\newcommand{\opleq}{\preceq}
\newcommand{\opgeq}{\succeq}

\newcommand{\eg}{{\it{e.g.}~}}
\newcommand{\ie}{{\it{i.e.}~}}

\newcommand{\ket}[1]{\vert #1 \rangle}
\newcommand{\bra}[1]{\langle #1 \vert}

\newcommand{\ketbra}[2]{|#1\rangle\langle#2|}

\definecolor{aqua1}{HTML}{4FC1E9}
\definecolor{aqua2}{HTML}{3BAFDA}
\definecolor{mint1}{HTML}{48CFAD}
\definecolor{mint2}{HTML}{37BC9B}
\definecolor{bittersweet1}{HTML}{FC6E51}
\definecolor{bittersweet2}{HTML}{E9573F}
\definecolor{lightgrey1}{HTML}{F5F7FA}
\definecolor{lightgrey2}{HTML}{E6E9ED}
\definecolor{grapefruit1}{HTML}{ED5565}
\definecolor{grapefruit2}{HTML}{DA4453}
\definecolor{sunflower1}{HTML}{FFCE54}
\definecolor{sunflower2}{HTML}{F6BB42}
\definecolor{bluejeans1}{HTML}{5D9CEC}
\definecolor{bluejeans2}{HTML}{4A89DC}
\definecolor{lavander1}{HTML}{AC92EC}
\definecolor{lavander2}{HTML}{967ADC}
\definecolor{pinkrose1}{HTML}{EC87C0}
\definecolor{pinkrose2}{HTML}{D770AD}
\definecolor{grass1}{HTML}{A0D468}
\definecolor{grass2}{HTML}{8CC152}
\definecolor{IoPpurple}{HTML}{A89CCE}
\definecolor{IoPdarkpurple}{HTML}{7E6FB0}

\definecolor{bsblue}{HTML}{0d6efd}
\definecolor{bsindigo}{HTML}{6610f2}
\definecolor{bspurple}{HTML}{6f42c1}
\definecolor{bspink}{HTML}{d63384}
\definecolor{bsred}{HTML}{dc3545}
\definecolor{bsorange}{HTML}{fd7e14}
\definecolor{bsyellow}{HTML}{ffc107}
\definecolor{bsgreen}{HTML}{28a745}
\definecolor{bsteal}{HTML}{20c997}
\definecolor{bscyan}{HTML}{17a2b8}

\colorlet{maincolor}{bspurple}
\colorlet{altcolor}{bsorange}
\colorlet{lightgrey}{black!5}

\colorlet{varcolour}{bsorange}
\colorlet{dualcolour}{bsred}
\colorlet{datacolour}{bsblue}
\colorlet{linkcolour}{bspurple}

\newcommand{\var}[1]{{\color{varcolour}#1}}
\newcommand{\dual}[1]{{\color{dualcolour}#1}}
\newcommand{\data}[1]{{\color{datacolour}#1}}

\newcommand\alert[1]{\textit{#1}}
\newcommand\lesson[1]{\textcolor{maincolor}{\textit{#1}}}

\title{Semidefinite Programming in Quantum Information Science}
\author{Paul Skrzypczyk and Daniel Cavalcanti}
\date{March 2023}

\begin{document}
\frontmatter
\maketitle
\tableofcontents

\chapter*{Preface}\addcontentsline{toc}{chapter}{Preface} 
\label{Preface}
We can safely say that learning semidefinite programming was a turning point in both of our careers. It all started in 2013, when we both joined Prof. Antonio Ac\'in's group at ICFO--The Institute of Photonic Sciences (Barcelona) to conduct our postdoctoral research. At that time we became interested in the problem of characterising quantum steering, a form of quantum correlations that were first noticed by E. Schr\"odinger in response to Einstein, Podolsky and Rosen's famous 1935 paper. We suspected that there was more to discover about quantum steering, and the role it can play in quantum information science.  It was then, during the traditional Quantum Information Conference `13, held in the beautiful mountain village of Benasque (Spain), that Miguel Navascu\'es (now a group leader at the University of Vienna) mentioned that steering could be studied using semidefinite programming. We immediately started digging into this idea and, a few months later, we published our first paper (also in collaboration with Miguel) showing how to use semidefinite programming in the quantification of steering. Since then, we have both been using semidefinite programming directly to achieve results in quantum information and foundations, or indirectly to test assumptions, to the point that using this technique has become an obsession. In fact, this is the main reason why we decided to write this book: semidefinite programming has helped us tremendously, and we are sure that it will help other researchers too.

We also have to mention the push we got from the VII Paraty Quantum Information School in 2019, where one of us was invited to give a mini-course on the topic of this book. During this event we saw a big interest from students on the topic, and realised there was a lack of a concise and coherent source of information about it, that could appeal to theorist or experimentalist, mathematician or physicist, and all those in between.

Because of this, we would like to thank Antonio Ac\'in for providing us with a relaxed and inspiring collaborative environment in his group. We would also like to thank Miguel Navascu\'es, and everyone involved in the organisation of the Benasque Quantum Information Conference as well as the organisers and students of the VII Paraty Quantum Information School. We also thank our numerous colleagues who, for so many times, have helped us to better understand semidefinite programming. In particular Marco T. Quintino, Dennis Rosset, Jean D. Bancal, Valerio Scarani, Marco Piani, Ashley Montanaro, Tony Short, and Sandu Popescu. Finally, we would like to thank IOP Publishing, and in particular John Navas, for giving us the opportunity to write this book. 

\chapter*{About the authors}\addcontentsline{toc}{chapter}{About the authors} 

Paul Skrzypczyk is currently an Associate Professor, Royal Society University Research Fellow, and CIFAR Azrieli Global Scholar, at the University of Bristol in the School of Physics. He obtained his PhD in Theoretical Physics from the University of Bristol in 2011, under the supervision of Professor Sandu Popescu, with his PhD studies focusing on quantum nonlocality and quantum thermodynamics. He carried out postdoctoral research at the University of Cambridge, and ICFO--The Institute for Photonic Sciences, before returning to Bristol in 2015. In 2016 he was awarded a Royal Society University Research Fellowship, and became a lecturer in 2018 and an Associate Professor in 2022. Paul's research interest span many areas of quantum information and quantum foundations, ranging from quantum nonlocal effects, such as Bell nonlocality, quantum steering and quantum teleportation, to quantum measurements and measurement incompatibility, to quantum thermodynamics. Convex geometry, semidefinite programming and convex optimisation have been the primary mathematical tools used in his research over the years.

Daniel Cavalcanti is currently a senior researcher at Algorithmiq Ltd. He obtained a PhD in Theoretical Physics at ICFO--Institute of Photonic Sciences in 2008 on the topics of entanglement and characterisation of quantum correlations. After a short postdoc at ICFO in 2009, he joined the Centre for Quantum Technology (Singapore) in 2010, first as a postdoc in Professor Valerio Scarani's group, and soon as an independent researcher. In 2013 he returned to ICFO, where he stayed until 2021 on a Ram\'on y Cajal grant. Daniel's research has focused on quantum foundations, quantum correlations, quantum communication and, more recently, quantum computation. Daniel also holds a master's degree in graphic design and runs Bitflow, a graphic design studio dedicated to science and technology related projects.

\chapter*{About this book}\addcontentsline{toc}{chapter}{About this book} 

Semidefinite programming (SDP\footnote{Throughout this book we will use the acronym SDP to denote both \emph{semidefinite programming} and \emph{semidefinite program}.}) is a type of optimization problem with vast applications in physics, engineering, combinatorial optimization, and many other fields. In this book we are interested in applications of semidefinite programming in the field of quantum quantum information. It turns out to be pretty natural that many problems in quantum mechanics and quantum information can be cast as SDPs, because the mathematical description of quantum states and quantum measurements are in terms of positive semidefinite operators fit naturally in the SDP framework. Thus, although here we will mostly focus on quantum information, the present book can also be relevant when addressing other problems in quantum physics. 

From a practical point of view, there are two main reasons to study SDPs. First, once a problem is recognised as an SDP, there is a theoretical machinery that can be used to solve the problem or, at least, obtain bounds on its solution. Second, there are plenty of efficient computer algorithms and modelling languages for solving SDPs, such as \href{http://cvxr.com/cvx/}{CVX}, \href{https://cvxopt.org/}{CVXOPT}, \href{https://www.cvxpy.org/}{CVXPY}, \href{https://www.mosek.com/}{MOSEK} and \href{https://yalmip.github.io/}{YALMIP}, allowing us to numerically solve SDPs involving relatively big matrices with today's standard laptop computers. As a starting point for performing numerical calculations of SDPs we recommend in particular the \href{http://cvxr.com/cvx/}{CVX} and \href{https://www.cvxpy.org/}{CVXPY} modelling systems for constructing and solving convex optimization problems (of which SDPs are a subclass) that can be implemented in \href{https://www.mathworks.com}{Matlab} and \href{https://www.python.org/}{Python} respectively. This book will focus exclusively on the theoretical machinery of SDPs. However, to accompany it, we have set up a repository at \href{https://github.com/paulskrzypczyk/SDPBook}{https://github.com/paulskrzypczyk/SDPBook}, which contains a small number of example codes, covering some of the SDPs studied in this book.

The structure of this book is the following. It is split into two parts. Part I is devoted to the description of semidefinite programming,  focusing on its main aspects. It starts with a simpler sub-class of optimization problems called \alert{linear programs} (Chapter 1) and then moves on to the main description of SDPs (Chapter 2). In these two chapters we will also set up the notation, nomenclature and basic definitions and concepts that will be used throughout the book. Then, Part II is devoted to particular problems from across the realm of quantum information science that can be tackled with the help of semidefinite programming. The list of problems we consider here is by no means exhaustive. Nonetheless, it is important to explain a little about the choice of problems presented. First, we wanted to cover a wide range of problems in quantum information. In this regard, you will find chapters related to quantum states, measurements, entanglement and channels. Second, we have chosen problems specifically that allow us to introduce key methods and tools for transforming problems into SDPs, and for manipulating and analysing those SDPs. These methods and tools are those which we believe to be very useful for a reader who wants semidefinite programming to be a useful tool in their research or work. In this respect, we encourage the reader to go through all chapters, even if the particular topic of the chapter is not so relevant for them; the way the topic is studied should hopefully teach them something new and interesting about semidefinite programming beyond the topic itself. In any case, we list the most important messages of each chapter in a section called \emph{Concluding Remarks}.

The list of problems we cover are as follows:
\begin{itemize}
	\item Chapter 3 addresses one of most basic problems in quantum physics, how to determine the properties of quantum states produced by an uncharacterised source. There are numerous variants of this problem, the most well-known being quantum state tomography, that is, estimating exactly which state the source is emitting. In this chapter we will review this and other variants of state estimation, including the so-called quantum marginal problem. This will allow us to study various aspects of SDPs, including how to obtain certificates of feasibility and the prevalent notion of a `witness'. 
	
	\item In Chapter 4, we shift focus to quantum measurements. We will first discuss the problem of estimating which measurement a given measuring device is performing.  We then  study quantum state discrimination, the problem of determining -- by performing a measurement --  which state, out of a set of possibilities, a given source is producing. We show how to gain new insight into this problem by using the key concept of SDP duality, and how the tool of complementary slackness can be used to derive optimality conditions. 
	
	\item We then move on to the problem of characterising quantum entanglement in Chapter 5. Entanglement is nowadays seen as the main resource behind many quantum information tasks, such as quantum teleportation, computation and cryptography. In this chapter we will see how SDPs can be used to detect and quantify entanglement. We will also see how we can use a sequence -- or hierarchy -- of semidefinite programs to obtain approximations to the set of separable quantum states, and therefore bounds on many quantities of interest. 
	
	\item In Chapter 6 we will study one of the most basic and interesting properties of quantum mechanics: measurement incompatibility. We will see how this problem can be solved by semidefinite programming, and use duality to uncover an unexpected link to quantum nonlocality. 
	
	\item Finally, in Chapter 7 we will study quantum channels. We will see how the so-called Choi-Jamio\l kowski isomorphism -- which links quantum channels with quantum states -- allows for semidefinite programming techniques to be applied to quantum channels. We will then see how we can estimate an uncharacterised channel, calculate the diamond norm (which characterises the operational distinguishability between channels) and finally how to use duality to uncover a link between a channel optimisation problem and quantum entropies. 
\end{itemize}

At the end of each chapter we also suggest a small amount of additional material on each topic (in the section \emph{Further Reading}). It is important to note that the list of references is by no means exhaustive. Our aim is simply to provide the curious reader with further resources -- in the form of relevant textbooks or review articles -- to complement the present text. In this connection, let us stress that there are several texts on semidefinite programming that can be used to complement this book in general. In particular, the book \href{https://web.stanford.edu/~boyd/cvxbook/}{\emph{Convex Optimization}} by S. Boyd and L. Vandenberghe is a must-have reference on convex optimization, which includes and goes beyond the theory of SDPs, and provides the reader with a comprehensive amount of information on the subject (and is freely available!). More related to quantum information are the textbook \href{https://cs.uwaterloo.ca/~watrous/TQI/}{\textit{The Theory of Quantum Information}} and lecture notes \href{https://cs.uwaterloo.ca/~watrous/TQI-notes/}{\textit{Theory of Quantum Information}} by J. Watrous and the lecture notes \href{https://sites.google.com/site/jamiesikora/teaching/semidefinite-programming-quantum-information}{\textit{Semidefinite Programming \& Quantum Information}} by J. Sikora and A. Varvitsiotis, all of which are freely available online. 

Finally, this book is very much just an introduction to the topic of semidefinite programming. There are many more advanced topics that we chose not to cover here. Our goal was to provide a solid foundation to the key aspects of the theory, and to demonstrate their widespread applicability in quantum information science. We hope that using this book as a foundation, those readers who choose to do so, will be well placed to go on and master more advanced aspects of semidefinite programming -- and more generally convex optimisation -- and put them to good use in whichever direction they see fit. 

\mainmatter

\setcounter{chapter}{2}

\chapter{Quantum states}\label{ch:q states}
\setcounter{ex-num}{0}
In this chapter we begin the second -- and main -- part of this book, where we explore the use of semidefinite programming in the context of quantum information science. As will be seen, semidefinite programming proves to be a versatile and powerful tool, which finds widespread and varied applications. 

One of the first reasons why semidefinite programming is so useful in quantum information is because -- mathematically -- quantum states are positive semidefinite operators, $\rho\opgeq 0$, which are normalised, $\tr(\rho)=1$. Being a linear operator inequality and an equality constraint respectively, these conditions can be used as constraints inside SDPs, so that many  problems involving optimisation over quantum states can be cast as SDPs. In this chapter we will describe some of these problems. We will discuss problems related to \alert{quantum state estimation}, which refers to the question of estimating the state that a source produces given the outcome statistics of a set of measurements. In the case this set of measurements is tomographically complete, this is equivalent to quantum state tomography -- characterising the state emerging from a source. Sometimes, however, the information obtained experimentally is not enough to single out a unique quantum state. With the help of SDPs we can nevertheless characterise the set of states compatible with a given set of experimental observations, and estimate additional (unmeasured) properties of quantum states. 

Altogether, these topics will allow us to see how problems involving quantum states can be cast as SDPs, and to use various aspects of the theory of SDPs presented in the previous chapters, in order to gain important insights about these problems.

We will end this chapter with a brief section on the \alert{quantum marginal problem}, which can be viewed as a generalisation of quantum state estimation. In this problem, we consider \alert{multipartite} systems, and are given complete information only about their \alert{subsystems}. The goal is then to determine the global state, or properties of it. As will be  seen, the tools of semidefinite programming can also be used to solve this problem, and leads to a simple way to see when certain sets of marginals are inconsistent with each other. 

\section{Quantum state estimation}

Suppose an experimentalist comes to you and says 
\begin{quote}
	\emph{I have a source of spin-1/2 particles (\ie qubits) in my laboratory, that I have measured along two different spin directions $\hat{x}$ and $\hat{y}$, obtaining the following expectation values: $\langle\sigma_x\rangle = 0.9$ and $\langle\sigma_y\rangle=0.5$,}
\end{quote}
where $\sigma_x$, $\sigma_y$ and $\sigma_z$ are the three Pauli operators. Is there any way to know if they are telling the truth or not? More precisely, is it possible to determine whether what they claim to have observed is consistent with the predictions of quantum mechanics?  The simple situation above can indeed be checked through basic algebraic means, using the properties of the measurements used. But in more complicated and sophisticated situations, involving, \eg many measurements with no symmetry relations between them, or high-dimensional quantum system, the task becomes more intricate.  

However, this problem can in fact be cast as a simple feasibility SDP. Namely, suppose that we are given the measurement results for a set of $N$ observables $\data{M_1}, \ldots,\data{M_N}$. Let us assume that we are only told the \alert{average} values of each measurement,  $\langle \data{M_1}\rangle=\data{m_1},\ldots,\langle \data{M_N}\rangle = \data{m_N}$. Our goal is to check if these reported values are compatible with some quantum state $\var{\rho}$. This amounts to solving the following feasibility SDP:
\begin{subequations}\label{eq: State Estimation}
	\begin{align}
		\text{find} &\quad \var{\rho} \\
		\text{subject to} &\quad \tr(\data{M_x} \var{\rho}) = \data{m_x}&x = 1,\ldots, N \label{e:QSE eq}\\
		&\quad\tr(\var{\rho}) = 1,\\
		&\quad \var{\rho} \opgeq 0. 
	\end{align}
\end{subequations}
There are three possibilities that can arise. 
\begin{enumerate}
	\item \textbf{Unique solution.} The set of observables $\{\data{M_x}\}$ we consider might have the property of being \alert{tomographically complete}. This means that knowing the results for every measurement in the set is sufficient to uniquely determine the state. Mathematically, the set must have a sufficient number of linearly independent measurements in order to determine all of the matrix elements of the density operator $\var{\rho}$. In this case, the feasible set of the SDP is a unique density operator $\mathcal{F} = \var{\rho^*}$.
	\item \textbf{No solution.} The other extreme case is when there is \alert{no solution}, \ie there is no quantum state $\var{\rho}$ which is able to produce the desired list of expectation values (and the experimentalist was lying or mistaken). In this case, the SDP is \alert{infeasible}, and we see that the feasible set must be empty, $\mathcal{F} = \emptyset$. We will study this in detail in Sec.~\ref{s:cert infeas}, where it is shown that the dual SDP can be used to provide us with a certificate, which will convince us that there is no quantum state producing the desired set of expectation values. 
	\item \textbf{Multiple solutions}. The situation in between the above two cases, is that there are infinitely many states $\var{\rho}$ which produce the given set of expectation values. This means that the set of measurements is not tomographically complete, such that they do not uniquely determine all of the matrix elements of the density operator. On the other hand, the expectation values are consistent with each other. In terms of the SDP, here the feasible set $\mathcal{F}$ will be non-empty, and contain more than a single point.
\end{enumerate}

\subsection{Trace distance estimation}\label{s:qse td}
The above shows that quantum state estimation can be cast as an SDP. We can however use semidefinite programming to dig much deeper into the problem of state estimation. As a first example, imagine that not only do we have the set of measurement results, but also a \alert{target state}, that we believe was prepared, and which was measured to produce the observed statistics. If the statistics are incompatible with this target state we might be interested in understanding \alert{how close} the prepared state was to the target state.

In this section we see that this question can be answered using semidefinite programming. In particular, we will show that it is possible to find the \alert{best-case} trace-distance between any state consistent with the measurement results and some target state. 

The key ingredient is to use the primal SDP formulation for the trace norm, as given in (2.8), which we will see can be harnessed here for our needs. 

Recall that for a pair of quantum states the trace distance is given by
\begin{align}\label{e:trace distance}
	T(\rho,\sigma) &= \frac{1}{2}\| \rho - \sigma \|_1.
\end{align}
This is a non-linear function of $\rho$ and $\sigma$, and so cannot be used directly as either the objective function or as a constraint inside an SDP. However, as we saw in Example 2.3 in the previous chapter, it can nevertheless be possible to recast problems which appear non-linear as SDPs. We will see here that trace distance estimation is indeed such an problem where the non-linearity is not a road block.  The first thing to note is that although non-linear, the trace distance is nevertheless a convex function of either state, such that
\begin{equation}
	T(q \rho_0 + (1-q) \rho_1,\sigma) \leq q T(\rho_0,\sigma) + (1-q)T(\rho_1,\sigma), 
\end{equation}
and similarly for the second state. 

There are two possibilities that are of potential interested -- the \alert{best-case} and the \alert{worst-case}. In the former, we would want to find the \alert{closest} state to our target state consistent with the measurement results. On the other hand, in the latter case, we would seek to find the state which is furthest away from our target state and still consistent with the results. 

The fact that the trace distance is convex means that we will not be able to analyse the \alert{worst-case}, but only the best-case. Why is this? This is in fact an important lesson about convex optimisation -- more general than just LPs and SDPs. As the following sketch illustrates, \alert{convex} objective functions are naturally associated with  \alert{minimisation} problems and \alert{concave} objective functions are naturally associated with \alert{maximisation} problems: maximising a convex function we are led to the extremes of the function -- to the boundary of the domain. This is a fundamentally different type of optimisation problem compared with finding the optima, which naturally occur at \alert{stationary points}. 

\begin{figure}[h!]
	\centering
	\includegraphics[width=0.7\linewidth]{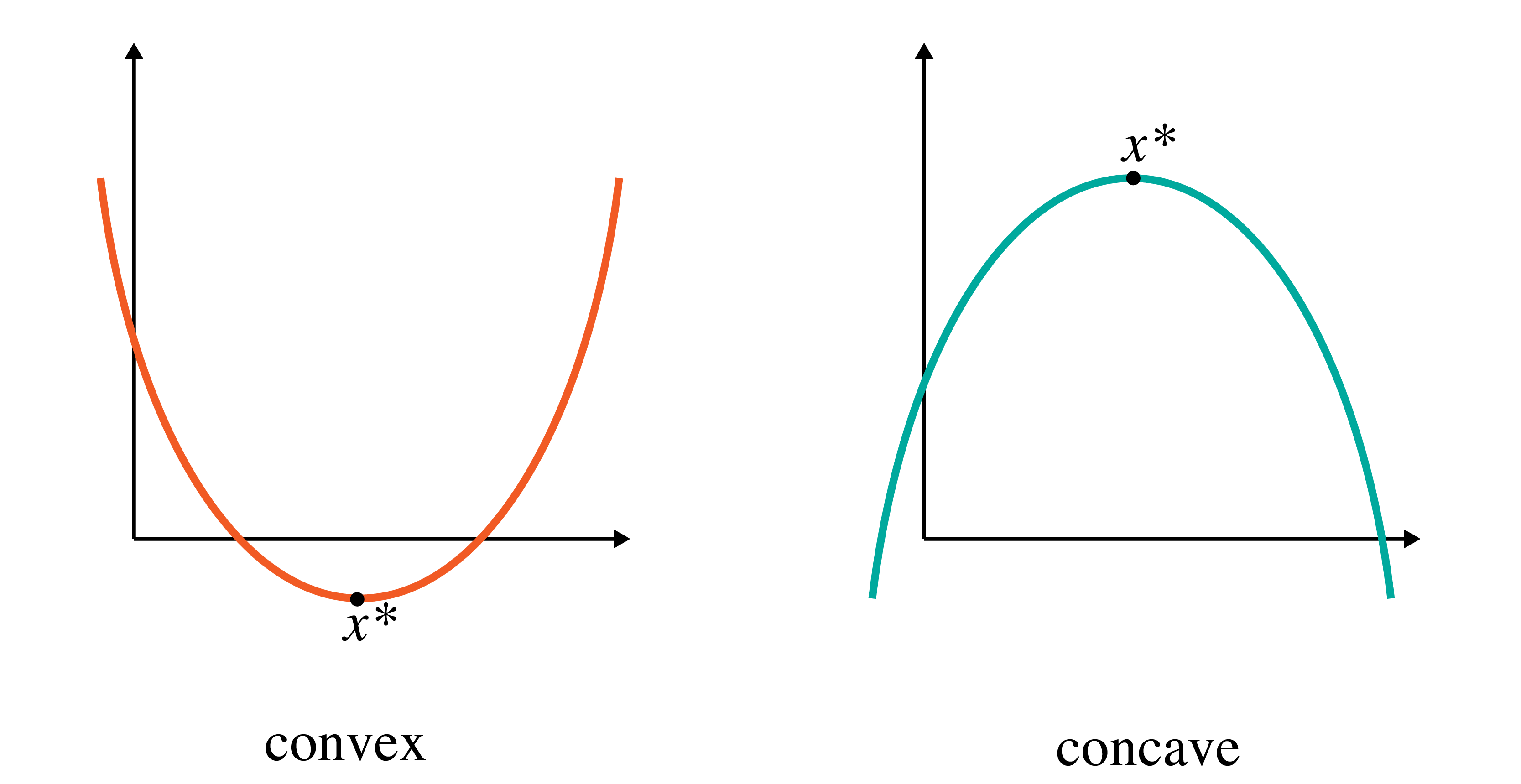}
\end{figure} 

Since the trace distance is convex, we can consider problems involving its minimisation, which here would correspond to finding the \alert{minimal distance} or \alert{closest state} to the target state.  In particular, the problem we can attempt to solve is
\begin{subequations}\label{e:QSE td orig}
	\begin{align}
		\text{minimise} &\quad \tfrac{1}{2}\|\var{\rho} - \data{\sigma}\|_1\\
		\text{subject to}&\quad \tr (\data{M_x} \var{\rho}) = \data{m_x}& &x=1,\ldots,N, \\
		&\quad \tr (\var{\rho}) = 1, \\
		&\quad \var{\rho} \opgeq 0.
	\end{align}
\end{subequations}
This is still not an SDP, but it can be turned into one by making use of Example 2.3,  in particular (2.8), which gave an SDP for the trace norm $\|A\|_1$ of an arbitrary Hermitian operator $A$. Here there are two additional complications: (i) Whereas previously $A$ was the \alert{data} of the problem, here $A = \var{\rho} - \data{\sigma}$ is instead a \alert{variable}. (ii) As stated above, we now want to \alert{minimise} the trace norm rather than just calculate it. Luckily both of these complications are harmless. First, in (2.8) the variable $\var{X}$ is never multiplied by the data $\data{A}$. Thus, although here $A = \var{\rho} - \data{\sigma}$ is now a variable, it doesn't actually stop the problem from being linear, and therefore from being an SDP. 

Second, since the SDP (2.8) for the trace norm is a minimisation, and we want to consider here the \alert{best case} -- itself corresponding to \alert{minimisation} -- we end up with a double minimisation. However, double minimisations are still convex optimisation problems, and in this case our problem \alert{remains an SDP}. Altogether, we arrive at the following SDP for finding the closest state in trace distance to a target state, given the measurement results:
\begin{subequations}\label{e:trace distance SDP data}
	\begin{align}
		\text{minimise} &\quad \tfrac{1}{2} \tr(\var{X})\\
		\text{subject to} &\quad -\var{X} \opleq \var{\rho} - \data{\sigma} \opleq \var{X},\\
		&\quad \tr (\data{M_x} \var{\rho}) = \data{m_x},& &x=1,\ldots,N \\
		&\quad \tr (\var{\rho}) = 1, \\
		&\quad \var{\rho} \opgeq 0.
	\end{align}
\end{subequations}
Incidentally, we can now understand mathematically why it is not possible to analyse the worst case, \ie to find the state which is as far as possible from the target state. In such a case, we would end up with almost the same problem as \eqref{e:QSE td orig}, except the objective would now be to maximise, rather than minimise. However, we would then not be able to end up at a form like \eqref{e:trace distance SDP data}, since there it was possible to combine the two minimisations -- one over $\var{\rho}$ and one over $\var{X}$ -- into a single minimisation, and obtain an SDP. To evaluate the worst case, we would need to maximise over $\var{\rho}$ and minimise over $\var{X}$, and these \alert{cannot be combined} into a single optimisation. This reflects the fact that to find the worst case, as already realised, would require maximising a convex function, which is a fundamentally different type of problem. 

\subsection{Fidelity estimation}\label{s:qse fid}

We now turn our attention to a second notion of closeness between states -- namely the fidelity. We will start by specialising to the problem of determining how close $\var{\rho}$ is to a \alert{pure} quantum state $\data{\sigma}=\data{\ketbra{\psi}{\psi}}$. As will be seen, this is more straightforward than the  general problem of estimating the fidelity to a \alert{mixed} state. In particular, the fidelity between a state $\rho$ (which can be either mixed or pure) and a pure state $\sigma = \ket{\psi}\bra{\psi}$ is 
\begin{equation}
	F(\rho,\sigma) = \bra{\psi}\rho\ket{\psi} = \tr(\ket{\psi}\bra{\psi}\rho),
\end{equation}
which, notably, is a \alert{linear} function of $\rho$. We can thus use this as the objective function directly, and arrive at the following SDP to find the closest state in fidelity to a target state, \ie the state with \alert{maximum fidelity} with the target state $\data{\sigma}=\data{\ketbra{\psi}{\psi}}$,
\begin{subequations}\label{eq: max eig}
	\begin{align}
		\text{maximise} &\quad  \tr(\data{\ket{\psi}\bra{\psi}}\var{\rho}) \\
		\text{subject to} 
		&\quad \tr (\data{M_x} \var{\rho}) = \data{m_x},& &x=1,\ldots,N \\
		&\quad \tr (\var{\rho}) = 1, \\
		&\quad \var{\rho} \opgeq 0.
	\end{align}
\end{subequations}
This SDP again allows us to understand the \alert{best-case}, \ie the largest fidelity among all possible states that are compatible with the observed data. 

Interesting, since the fidelity is linear, we can now consider the associated \alert{minimisation} problem, \ie to replace the maximisation in \eqref{eq: max eig} by a minimisation. In this case, we then find the \alert{worst-case} fidelity, \ie the smallest fidelity among all possible states consistent with the measurement results. 

This is somewhat surprising, since we saw for the trace distance that it isn't possible to obtain the worst-case state. The key difference here is that the fidelity to a pure state is linear, meaning it is both \alert{convex and concave}, and so we can in fact also solve the worst-case problem too. Unfortunately, as will be seen below, this is a special property of the fidelity with pure states. In the more general case, of fidelity with a mixed target state, it is again only possible to solve the best-case scenario. 

Moving on to the more general problem of estimating the fidelity between two mixed states, the general expression for the fidelity is given by
\begin{subequations}
	\begin{align}
		F(\rho, \sigma)  &= \left\|\sqrt{\rho}\sqrt{\sigma}\right\|^2_1,\\
		&= \left[\tr\left(\sqrt{\sqrt{\sigma}\rho\sqrt{\sigma}}\right)\right]^2,
	\end{align}
\end{subequations}
which is now a non-linear function of both $\rho$ and $\sigma$. This means that, just as with the trace distance, we can no longer directly use this as the objective function in an SDP. Although non-linear, the fidelity  is nevertheless a \alert{concave} function (in either state), meaning that
\begin{equation}
	F(q \rho_0 + (1-q) \rho_1,\sigma) \geq q F(\rho_0,\sigma) + (1-q)F(\rho_1,\sigma), 
\end{equation}
for $q \in [0,1]$, and similarly for $\sigma$. This says that the fidelity between the average state $\rho' = q \rho_0 + (1-q) \rho_1$ and $\sigma$ is never smaller than the average of the fidelities between $\rho_0$ and $\sigma$ and $\rho_1$ and $\sigma$. 

The fact that the fidelity is concave means -- once again -- that it will only be possible to estimate the \alert{best-case} fidelity, since we will be able to consider \alert{maximising} the fidelity (and now a larger fidelity means a closer state), given the constraints arising from the measurements result. 

The difficulty that must still be overcome is to understand how we can evaluate the fidelity objective function using semidefinite programming. We will return to this  topic in more detail in the Advanced topics Sec.~\ref{s:fidelity SDP}, but just as with the trace distance, even though the fidelity is a non-linear function, it can nevertheless be expressed as an SDP by introducing auxiliary variables.  As shown in Sec.~\ref{s:fidelity SDP}, the square-root-fidelity is in fact given by the following SDP:
\begin{subequations}\label{e:F2 gen}\begin{align}
		\sqrt{	F(\data{\rho},\data{\sigma})} = \text{maximise}&\quad \tr(\var{Y}) \\
		\text{subject to}&\quad \begin{pmatrix}\data{\rho} & \var{Y} + i \var{Z}\\ \var{Y} - i \var{Z} & \data{\sigma}\end{pmatrix} \opgeq 0. \label{e:F2 cons}
\end{align}\end{subequations}
We have written the constraint \eqref{e:F2 cons} in a relatively intuitive way, arranging the variables $\var{Y}$ and $\var{Z}$ into a \alert{block matrix}, along with the data of the problem -- the quantum states $\data{\rho}$ and $\data{\sigma}$. It is important to appreciate that this inequality constraint is equivalent to the standard form as in (2.1c), namely of the form $\Gamma(\var{X}) \opleq C$. In particular, since we have a pair of variables $\var{Y}$ and $\var{Z}$, we have a map $\Gamma(\var{Y},\var{Z})$, which can be written
\begin{equation}\label{e:F2 cons dirac}
	\Gamma(\var{Y},\var{Z}) = \ket{0}\bra{0} \otimes \data{\rho} + \ket{0}\bra{1}\otimes (\var{Y} + i \var{Z}) +  \ket{1}\bra{0}\otimes (\var{Y} - i \var{Z}) + \ket{1}\bra{1} \otimes \data{\sigma}
\end{equation} 
with a similar technique applicable for any block matrix. As can be seen, the block matrix form of \eqref{e:F2 cons} is much easier to read than the form of \eqref{e:F2 cons dirac}, and hence we will use the block form whenever appropriate to do so. 

Returning to the problem of estimating  the largest fidelity with a target state, we can now directly use \eqref{e:F2 gen} in order to recast the problem as an SDP. In particular, the following SDP is arrived at, which evaluates the best-case (square-root) fidelity between any state $\var{\rho}$ consistent with the data, and a target state $\data{\sigma}$:
\begin{subequations}\label{e:F2 optim}\begin{align}
		\text{maximise}&\quad \tr(\var{Y}) \\
		\text{subject to}&\quad \begin{pmatrix}\var{\rho} & \var{Y} + i \var{Z}\\ \var{Y} - i \var{Z} & \data{\sigma}\end{pmatrix} \opgeq 0,\\
		&\quad \tr (\data{M_x} \var{\rho}) = \data{m_x}& &x=1,\ldots,N, \\
		&\quad \tr (\var{\rho}) = 1, \\
		&\quad \var{\rho} \opgeq 0.
\end{align}\end{subequations}
Note that, compared with the SDP formulation of $\sqrt{F(\rho,\sigma)}$, we have relaxed $\rho$ from being \alert{input data} to being a \alert{variable}, constrained by the measurement results. Crucially, as can be seen, this problem remains \alert{linear}, and subsequently remains an SDP. We can also note that, just as for the trace distance, the above works since the SDP formulation of the fidelity was a \alert{maximisation} problem. By relaxing $\var{\rho}$, the best-case state, given the constraints, is now naturally found.

In principle, distances other than the trace distance and fidelity can also be considered in state estimation. As the above two examples have hopefully demonstrated, these distances need not be linear functions, but they must be expressible themselves as SDPs. 

\subsection{Finite statistics}\label{s:qse fs}

We now turn our attention to a different aspect of quantum state estimation, that of \alert{finite statistics}. In practice, due to the fact that every experiment can only handle a finite number of measurement rounds, it is rather unreasonable to assume that the expectation values $\data{m_1},\ldots,\data{m_N}$ are known exactly. In reality, these will have been estimated themselves, and will only be known up to some level of uncertainty. Let us therefore assume that the result of each measurement is only known to lie in some interval $[\data{m_x} - \data{\Delta_x}, \data{m_x} + \data{\Delta_x}]$. The problem of interest now is the feasibility problem of determining whether there is a quantum state $\var{\rho}$ which is consistent with all of these uncertain expectation values, 
\begin{subequations}\label{eq: State Estimation FS}
	\begin{align}
		\text{find} &\quad \var{\rho} \\
		\text{subject to} &\quad \data{m_x} - \data{\Delta_x} \leq \tr(\data{M_x} \var{\rho}) \leq \data{m_x} + \data{\Delta_x},& &x = 1,\ldots,N\label{e:QSE FS cons}\\
		&\quad\tr(\var{\rho}) = 1,\\
		&\quad \var{\rho} \opgeq 0. 
	\end{align}
\end{subequations}
where to aid presentation we have displayed \alert{both} inequality constraints related to the measurement intervals in a single line for each expectation value. Each equality constraint from the original problem \eqref{eq: State Estimation} is thus replaced by a pair of inequality constraints. As will be seen later, this will have an effect on the \alert{dual SDP} -- which will have twice as many dual variables compared to the dual of the original problem. 

From the perspective of the structure of the SDP, since there are no equality constraints in this problem, even if the set of measurements $\{\data{M_x}\}_x$ is tomographically complete, in general we would never expect to find a \alert{unique} solution, unlike before. This is rather natural, as we would expect to have some residual uncertainty about the state, given the uncertainty in the measurement results, whenever they are consistent. 

\subsection{Relaxing the feasibility problem}
In another direction, we return now to our original feasibility problem of quantum state estimation, as given in \eqref{eq: State Estimation}, and look at how it can be relaxed to an optimisation problem. There are a few of reasons for doing this. First, it will allow us to understand more \alert{quantitatively} inconsistent a set of measurement results are, whenever there is no state that can lead to them. Second, from a numerical perspective, it is useful to be able to transform feasibility problems into optimisation problems, which are generally much more stable to solve. Finally, having an optimisation form will allow us to use duality, which we will show to be useful in later sections. 

There are multiple ways in which the feasibility problem \eqref{eq: State Estimation} can be relaxed to an optimisation problem, each with their relative merits. Here we will adapt the approach from the previous section on finite statistics, in order to obtain a relaxation of the problem, which is \alert{always feasible}.

The key idea is to relax the equality constraints \eqref{e:QSE eq} -- to not demand that the observed expectation values are reproduced exactly, but allow them to be reproduced \alert{approximately}, in a very similar fashion to the above section. Whereas in the above the relaxation was coming from the uncertainty due to finite statistics, here the logic is different. Here we seek to find a set of measurement results which \alert{best approximates} the target results. If they can be reproduced exactly, then the original problem was feasible, and hence we obtain a quantum state $\var{\rho}$ that can reproduce the statistics. If on the other hand the problem is infeasible, it will be necessary to perturb the results in order for them to be producible by some quantum state. The crucial observation is that \alert{a sufficiently big perturbation of the results will always be producible by some quantum state}. This means that this relaxation will lead to an SDP which will be \alert{feasible by construction}. Finally, by minimising the size of the perturbation, the best approximation is found. 

It is interesting to note that this approach can be seem as complementary to that taken in Sections \ref{s:qse td} and \ref{s:qse fid} on trace distance and fidelity estimation. In particular, in those sections, in essence we sought to find the state which best approximated a target state. Here, on the other hand, we seek to find the set of measurement results which best approximates the target measurement results. 

A simple way to achieve the above relaxation is to place a \alert{uniform bound} on how much any single result can differ from the desired result. That is, we introduce a new variable $\var{\delta}$, and replace the equality constraints \eqref{e:QSE eq} with pairs of inequality constraints
\begin{equation}\label{e:QSE relax cons}
	\data{m_x}	-\var{\delta} \leq \tr(\data{M_x}\var{\rho})  \leq \data{m_x} + \var{\delta} \quad x = 1,\ldots, N.
\end{equation}
This enforces that for all measurements, $|\tr(\data{M_x}\var{\rho}) - \data{m_x}| \leq \var{\delta}$. The reason for writing this in the form \eqref{e:QSE relax cons} is that it in this form the constraint is manifestly linear in all of the variables. It is also worth noting that this is very similar to \eqref{e:QSE FS cons}, the difference being that previously the uncertainties $\data{\Delta_x}$ depended upon the measurement, and were input data to the problem, whereas here we simplify and consider only a single $\var{\delta}$, which is now constant and a variable of the problem. Putting everything together, we arrive at the following relaxed SDP for quantum state estimation:
\begin{subequations}		\label{eq: State Estimation optim}
	\begin{align}
		\text{minimise} &\quad \var{\delta} \\
		\text{subject to} &\quad \data{m_x} - \var{\delta} \leq \tr(\data{M_x} \var{\rho}) \leq \data{m_x} + \var{\delta}& &x = 1,\ldots,N, \label{e:QSE eq1}\\
		&\quad\tr(\var{\rho}) = 1 \label{e:QSE eq2},\\
		&\quad \var{\rho} \opgeq 0. \label{e:QSE relax}
	\end{align}
\end{subequations}
Any solution with $\var{\delta} = 0$ satisfies all of the constraints of the original problem \eqref{eq: State Estimation}. On the other hand, when $\var{\delta^*} > 0$, this signifies that the original problem \eqref{eq: State Estimation} is infeasible, and there is no quantum state able to reproduce the data. 

It is important to realise that $\var{\delta^*}$ provides us with quantitative information regarding how close the problem is to being feasible: if $\var{\delta^*}$ is small this means that there is a quantum state that is able to reproduce closely the measurement results; on the other hand, if $\var{\delta^*}$ is large, at least one of the measurement results needs to be significantly different from that observed, in order to be producible by some quantum state. 

\begin{exercises}
	\item By defining vectors $\data{\vec{m}}$ with components $\data{m_x}$ and $\var{\vec{m}'}$ with components $\var{m_x'} = \tr(\data{M_x}\var{\rho})$,  show that the SDP \eqref{eq: State Estimation optim} can also be written as
	\begin{align*}
		\text{minimise} &\quad \|\var{\vec{m}'} - \data{\vec{m}}\|_\infty \\
		\text{subject to} &\quad \var{m_x'} = \tr(\data{M_x}\var{\rho}),& &x = 1,\ldots,N, \\
		&\quad\tr(\var{\rho}) = 1 ,\\
		&\quad \var{\rho} \opgeq 0. 
	\end{align*}
	\lesson{This shows that this relaxation can be understood as minimising the $\ell_\infty$ distance between the target data $\data{\vec{m}}$, and any set of data that can be produced in quantum mechanics.}
	\item Write down the SDP relaxation of \eqref{eq: State Estimation} which minimises instead the $\ell_1$ distance $\frac{1}{2}\|\data{\vec{m}} - \var{\vec{m}'}\|_1$, between the target data $\data{\vec{m}}$, and data that can be produced in quantum mechanics $\var{\vec{m}'}$.  
\end{exercises}

\subsection{Certificate of infeasibility}\label{s:cert infeas}
We now consider the situation where we are given some experimental data which is inconsistent, such that there is no quantum state that could possibly lead to this data. An interesting question is: can we certify that this is the case without having to numerically solve the SDP? Is there an \alert{analytic} certificate that can be used to prove unequivocally that the SDP is infeasible? 

In this section it will be seen that duality allows us to provide such a simple \alert{certificate}, which will convince us whenever a set of measurement results is inconsistent. This idea is rather general, and can be widely applied to guarantee that an given SDP is infeasible.

Our starting point will be the relaxed problem \eqref{eq: State Estimation optim}. Recall that we attempt to minimise $\var{\delta}$, and if we are able to find a solution $\var{\delta} = 0$, then the associated state $\var{\rho}$ will be consistent with the measurement results. On the other hand, if $\var{\delta^*} > 0$, then the results are inconsistent. In what follows we will show how duality can be used to guaranteeing that $\var{\delta^*} > 0$ without having to solve the SDP. 
Recall that for a minimisation problem, the values that the dual objective function can take always lower bound the optimal value of the primal problem. Therefore, if we can find a set of dual variables such that the value of the dual objective function is strictly positive, then \alert{with certainty} the primal problem has $\var{\delta^*} > 0$. It is precisely a collection of dual variables that lead to a positive value of the dual objective function that will constitute our certificate. Let us now put this into practice. 

By introducing scalar dual variables $\dual{u_x}$ and $\dual{v_x}$ associated to the first and second inequality constraint respectively in \eqref{e:QSE eq1}, a scalar $\dual{z}$ associated to \eqref{e:QSE eq2} and an operator $\dual{W}$ associated to \eqref{e:QSE relax}, the Lagrangian is
\begin{subequations}
	\begin{align}
		\mathcal{L} &= \var{\delta} - \sum_{x=1}^N \dual{u_x}\left[\tr(\data{M_x}\var{\rho}) - \data{m_x} +\var{\delta} \right] - \sum_{x=1}^N \dual{v_x}\left[\data{m_x} +\var{\delta} - \tr(\data{M_x}\var{\rho})\right]+ \dual{z}[1-\tr(\var{\rho})] - \tr(\dual{W}\var{\rho}),\\
		&= \var{\delta}\left[1-\sum_{x=1}^N \left(\dual{u_x}+\dual{v_x}\right)\right] + \tr\left[\var{\rho}\left(\sum_{x=1}^N \left(\dual{v_x} - \dual{u_x}\right)\data{M_x} - \dual{z}\identity - \dual{W}\right)\right] + \dual{z} + \sum_{x=1}^N (\dual{u_x}-\dual{v_x})\data{m_x}.\label{e:L QSE 2}
	\end{align} 
\end{subequations}
Note that since the primal problem is a minimisation problem, the Lagrangian is constructed to be smaller than the value of the primal objective function for all primal feasible variables, and hence we take $\dual{u_x} \geq 0$, $\dual{v_x} \geq 0$ and $\dual{W} \opgeq 0$. Recall that this is the reason for the additional minus signs in the Lagrangian in the second, third and last term, which arise whenever we construct the Lagrangian for a minimisation problem, as shown in Exercise 2.11. We can additionally make the Lagrangian independent of the primal variables be ensuring the first and second brackets vanish in  \eqref{e:L QSE 2}. Maximising, to obtain the best lower bound, and solving for $\dual{W}$, which is seen to play the role of a slack variable, we arrive at the dual formulation
\begin{subequations}
	\begin{align}
		\text{maximise}&\quad \dual{z} + \sum_{x=1}^N (\dual{u_x}-\dual{v_x})\data{m_x} \\
		\text{subject to}&\quad \dual{z}\identity + \sum_{x=1}^N (\dual{u_x}-\dual{v_x})\data{M_x} \opleq 0, \label{e:QSE dual ineq}\\
		&\quad \sum_{x=1}^N (\dual{u_x} + \dual{v_x}) = 1,\label{e:QSE dual norm1} \\
		&\quad \dual{u_x} \geq 0,\quad \dual{v_x} \geq 0,& &x = 1,\ldots, N.\label{e:QSE dual norm2}
	\end{align}
\end{subequations}
Let us now analyse this dual a little. The first thing to notice is that the constraints \eqref{e:QSE dual norm1} and \eqref{e:QSE dual norm2}, together with the fact that everywhere else only the combination $\dual{u_x} - \dual{v_x}$ appears, shows that the problem can be interpreted as one of optimising over a single dual variable $\dual{\vec{t}} = \dual{\vec{u}} - \dual{\vec{v}}$, with components $\dual{t_x} = \dual{u_x} - \dual{v_x}$, such that $\|\dual{\vec{t}}\|_1 \leq 1$. This can be seen from the definition of the $\ell_1$ unit ball from (1.62). Thus, we can write this down in a simpler form:
\begin{subequations}\label{e:dual SDP state est}
	\begin{align}
		\text{maximise}&\quad \dual{z} + \dual{\vec{t}}\cdot \data{\vec{m}}\label{e:QSE dual obj} \\
		\text{subject to}&\quad \dual{z}\identity + \dual{\vec{t}}\cdot \data{\vec{M}} \opleq 0, \label{e:QSE dual ineq 2}\\
		&\quad \|\dual{\vec{t}}\|_1 \leq 1, \label{e:QSE dual 1norm}
	\end{align}
\end{subequations}
where we have introduced an \alert{operator vector} $\data{\vec{M}} = (\data{M_1},\ldots,\data{M_N})$, the components of which are the observables, and, as above, $\data{\vec{m}} = (\data{m_1},\ldots,\data{m_N})$ is the vector whose components are the measurement results.  

How does this serve as a certificate for the fact that the measurement results $\data{\vec{m}}$ were inconsistent? Let us assume that we have found a set of dual feasible points -- \ie a set of dual variables $(\dual{z},\dual{\vec{t}})$ satisfying the constraints -- such that $\beta = \dual{z} + \dual{\vec{t}}\cdot \data{\vec{m}} > 0$. Now, consider the constraint \eqref{e:QSE dual ineq 2}, and multiply both sides by an arbitrary quantum state $\rho$ and take the trace. We see that
\begin{equation}
	\tr\left[\rho\left(\dual{z}\identity + \dual{\vec{t}}\cdot\data{\vec{M}}\right)\right]   = \dual{z} + \sum_{x=1}^N \dual{t_x}\tr\left(\rho\data{M_x}\right)\leq 0.
\end{equation}
The first equality follows from the linearity of the trace, while the second follows from the fact that for two operators $A \opgeq 0$ and $B \opleq 0$, we have $\tr(AB) \leq 0$. What this shows is that the operator inequality \eqref{e:QSE dual ineq 2} guarantees that any set of measurement results $\vec{m}' = (\tr(\rho \data{M_1}),\ldots, \tr(\rho \data{M_N}))$ that can arise by performing measurements on a quantum state, will never lead to a positive value of $\beta$. This shows that the claim that the purported results $\data{\vec{m}}$ were observed, must have been false. The dual problem is thus seen to look for a carefully chosen set of dual variables $(\dual{z},\dual{\vec{t}})$ such that on the one hand, the purported results lead to a positive value of the dual objective function, while simultaneously demanding that all valid measurement results lead to a negative value. In the next section we will explore the geometry of this, but as a preview, what we have just observed can be thought of as a \alert{separation} between what is allowed and what is not allowed, and is a common geometrical feature of certificates of this type. 

Why is this better than solving the primal SDP in \eqref{eq: State Estimation}? There are a number of advantages. First, given the dual variables, no optimisation is required in order to check whether the following two conditions hold: (i) $\beta = \dual{z} + \dual{\vec{t}}\cdot \data{\vec{m}} > 0$, and (ii) all of the eigenvalues of $\dual{W} = \dual{z}\identity + \dual{\vec{t}}\cdot \data{\vec{M}}$ are negative. Second, this certificate can in principle even be checked analytically, that is, we can analytically evaluate $\beta$ and find the eigenvalues of $\dual{W}$, and thus verify analytically that all of the constraints of the dual are satisfied. This is in contrast to only having a numerical solution for the primal problem. Finally, although this certificate is designed for a given set of inconsistent data $\data{\vec{m}}$, it can be used more universally, to check the inconsistency of \alert{any} data. In particular, although it may fail to identify inconsistent data, if any set of data leads to a strictly positive value of $\beta$, then it is guaranteed to be inconsistent. We exemplify this in the following example:

\begin{example}[Equatorial plane of the Bloch sphere]\label{ex:bloch equator}
	In this example, we will show how the above dual SDP \eqref{e:dual SDP state est} can in fact be used as a method to derive the equator of the Bloch sphere. That is any qubit, written in the form $\rho = (\identity + \vec{r} \cdot \vec{\sigma})/2$, with $\vec{\sigma} = (\sigma_x,\sigma_y,\sigma_z)$ the vector of Pauli operators and $\vec{r}$ the Bloch vector with components $r_i = \tr(\sigma_i \rho)$, will be a valid density operator only if $r_x^2 + r_y^2 \leq 1$. 
	
	In order to obtain this result, assume that two Pauli measurements are performed, $\data{\vec{M}} = (\data{\sigma_x},\data{\sigma_y})$. We will use \eqref{e:dual SDP state est} to find the allowed measurement results $\data{\vec{m}} = (\data{m_x},\data{m_y})$ that can arise -- and in so doing re-derive the equatorial plane of the Bloch sphere.
	
	From \eqref{e:QSE dual ineq 2}, we are seeking dual variables that satisfy
	\begin{equation}
		\dual{z}\identity + \dual{t_x} \data{\sigma_x} + \dual{t_z} \data{\sigma_z} \opleq 0\label{e:XZ 1}
	\end{equation}
	and from \eqref{e:QSE dual 1norm}, we furthermore must have $\|\dual{\vec{t}}\|_1 = |\dual{t_x}| + |\dual{t_z}| \leq 1$. Since any operator of the form $n_x \sigma_x + n_y \sigma_y$ has eigenvalues $\pm 1$ whenever $\|\vec{n}\|_2 = \sqrt{n_x^2 + n_y^2} = 1$, we see that the eigenvalues of the operator $\dual{z}\identity + \dual{t_x} \data{\sigma_x} + \dual{t_y} \data{\sigma_y}$ are
	\begin{equation}
		\lambda_{\pm} = \dual{z} \pm \| \dual{\vec{t}}\|_2,
	\end{equation}
	and so we will satisfy \eqref{e:XZ 1} whenever
	\begin{equation}
		\dual{z} + \| \dual{\vec{t}}\|_2 \leq 0.
	\end{equation}
	We can therefore take $\dual{z} = -\| \dual{\vec{t}}\|_2$, so that the inequality is saturated. With this choice of $\dual{z}$, the dual objective function is
	\begin{equation}
		-\| \dual{\vec{t}}\|_2 + \dual{\vec{t}}\cdot \data{\vec{m}}.
	\end{equation}
	Now, any set of measurement outcomes $\data{\vec{m}} = (\data{m_x},\data{m_y})$ for which this is strictly positive is inconsistent, since we have just arranged it such that this can never happen. That is, $\data{\vec{m}}$ is inconsistent if
	\begin{equation}
		\dual{\vec{t}}\cdot \data{\vec{m}} > \| \dual{\vec{t}}\|_2.
	\end{equation}
	Up to this point, we still haven't specified $\dual{\vec{t}}$, other than the requirement that $\|\dual{\vec{t}}\|_1 \leq 1$. Let us therefore choose 
	\begin{equation}
		\dual{\vec{t}} = \frac{\data{\vec{m}}}{\|\data{\vec{m}}\|_1}.
	\end{equation}
	This implies that $\|\dual{\vec{t}}\|_2 = \|\data{\vec{m}}\|_2/\|\data{\vec{m}}\|_1$, and so the data is inconsistent whenever
	\begin{equation}
		\|\data{\vec{m}}\|_2^2 > \|\data{\vec{m}}\|_2,
	\end{equation}
	\ie $\|\data{\vec{m}}\|_2 > 1$. In other words, consistent measurement outcomes must satisfy $\|\data{\vec{m}}\|_2 \leq 1$. Written out in full, this equates to $\data{m_x}^2 + \data{m_y}^2 \leq 1$. Since $\data{\vec{m}}$ is nothing but the components of the Bloch vector of a qubit in the equatorial plane, this re-derives the equatorial plane of the Bloch sphere. 
	
	Our course this result is well-known and can be obtained more directly, by finding the eigenvalues of a qubit density operator -- we did not need to resort to duality theory of SDPs to derive this result. However, the utility of going through this example is that in more complicated situations, involving for example non-Pauli measurements, or higher dimensions, this approach can be still be applied, and provides a general method that can be used to understand the limitations of the quantum state space, and the measurement statistics it leads to. 
	
	As an exercise below, this same procedure is carried out with all three Pauli measurements, from which the entire Bloch sphere of a qubit can similarly be recovered. 
\end{example}
\begin{exercises}
	\item Apply the same argument as in Exercise~\ref{ex:bloch equator}, but for data $\data{\vec{m}} = (\data{m_x},\data{m_y},\data{m_z})$ arising from the three Pauli measurements $\data{\vec{M}} = (\data{\sigma_x},\data{\sigma_y},\data{\sigma_z})$ to show that the data is inconsistent whenever $\|\data{\vec{m}}\|_2 > 1$. Explain why this re-derives the Bloch sphere. 
\end{exercises}

\subsection{Geometrical interpretation}
We finish our exploration of quantum state estimation by investigating how the above certificates of infeasibility can be understood geometrically. Understanding SDPs and their duality geometrically can be very powerful, and provides insight and intuition for how and why duality works. 

\begin{figure}[h!]
	\centering
	\includegraphics[width=0.8\linewidth]{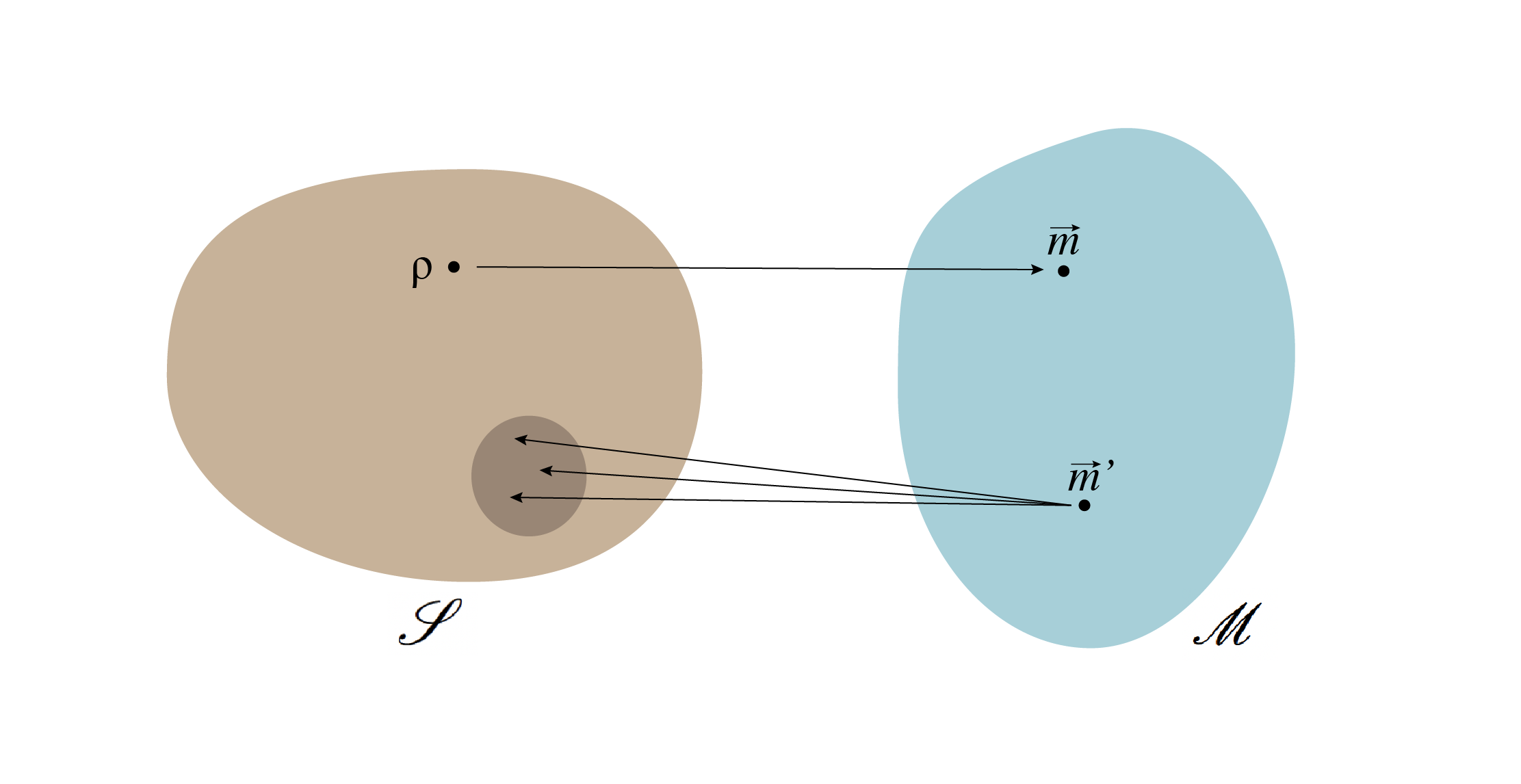}
	\caption{\alert{Mapping between state space and space of measurement results of a fixed set of measurements.} An illustration of the mapping of the quantum state space $\mathcal{S}$ into the set of measurement results $\mathcal{M}$, under the mapping $\rho \mapsto \data{\vec{m}} = (\tr(\data{M}_1\rho), \ldots, \tr(\data{M_N}\rho))$. For illustrative purposes both spaces are depicted as being 2-D, although in general they are both high-dimensional bodies of different dimension. Each quantum state $\rho$ is mapped to a single vector of measurement results $\vec{m}$. In general, the mapping can be many-to-one, with multiple quantum states being mapped to the same point in $\mathcal{M}$. This means that the inverse mapping, from a vector $\vec{m}'$ leads to a subset of the state space $\mathcal{S}$, as depicted. \label{f:mapping}}
\end{figure}

Our starting point is to think geometrically about the \alert{space of quantum states} and \alert{space of measurement data}. The space of all quantum states of a fixed, finite dimension $d$ is the set
\begin{equation}\label{e:statespace}
	\mathcal{S} = \{\rho \, | \rho \opgeq 0,\, \tr(\rho) = 1\}. 
\end{equation}
where $\rho$ is an operator acting on $\mathbb{C}^d$. This is a convex set, which we can always visualise as a subset of $\mathbb{R}^{d^2}$. For a fixed set of measurements $\data{\vec{M}}$, the set of all quantum states $\mathcal{S}$ gets mapped into a \alert{space of measurement results}, which is the image of $\mathcal{S}$ under the mapping
\begin{equation}\label{e:S to R mapping}
	\rho \mapsto \vec{m} = (\tr(\data{M_1}\rho), \ldots, \tr(\data{M_N}\rho))
\end{equation}
which can be represented as a subset of $\mathbb{R}^{N}$. We denote this image by $\mathcal{M}$. This mapping is illustrated in Fig.~\ref{f:mapping}. Because this is a linear map, the convexity of the state space is preserved, and the space of all measurement results is also a convex set -- See Exercise~\ref{exc:convex set}. 

We can understand many features of the state estimation problem from this geometrical perspective. First, note that the estimation problem is an \alert{inverse problem}: we are given a point in the measurement space, and ask for the quantum state(s) $\var{\rho}$ under the inverse mapping. When the set of measurements is not tomographically complete, then multiple states lead to the same set of measurement results. This shows that there is a \alert{loss of information}. As such, there is not a unique solution to the inverse problem. In this case, the dimension of the measurement space will be smaller than the dimension of the state space. On the other hand, when the measurements are tomographically complete, there is no loss of information and a one-to-one mapping between the two spaces. This implies that the dimension of the measurement space must be equal to the dimension of the state space.

\begin{figure}[h!]
	\centering
	\includegraphics[width=0.8\linewidth]{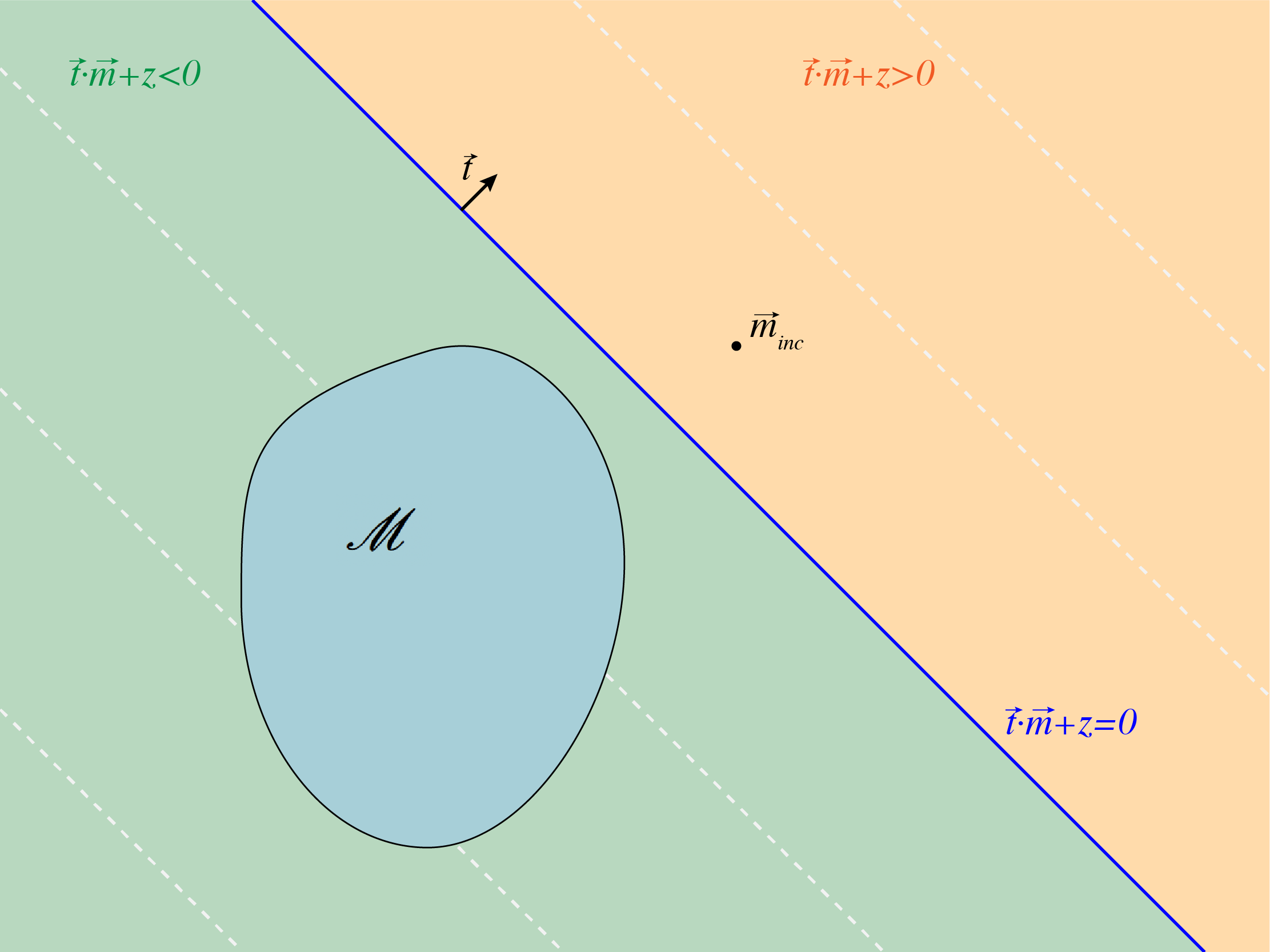}
	\caption{\alert{Geometric interpretation of certificate of infeasibility.} An illustrative example in 2-D of how the certificate of infeasibility can be viewed as a \alert{separating hyperplane} where $\dual{z} + \vec{\dual{t}}\cdot \data{\vec{m}} = 0$, which separates $\mathcal{M}$ -- the image of $\mathcal{S}$ (blue set) -- from the supposed set of measurement results $\data{\vec{m}_{\mathrm{inc}}}$.  The vector $\dual{\vec{t}}$ specifies a direction, and $\dual{z}$ specifies an offset, such that the image of $\mathcal{S}$ lies in the (green) region where $\dual{z} + \vec{\dual{t}}\cdot \data{\vec{m}} < 0$ and $\vec{m}_{\mathrm{inc}}$ lies in the orange region, where $\dual{z} + \vec{\dual{t}}\cdot \data{\vec{m}} > 0$. Re-arranging, we can view $-\dual{z}$ as the `bound' on the linear function $\vec{\dual{t}}\cdot \data{\vec{m}}$ that separates the two regions. \label{f:cert infeas} }
\end{figure}

Most interestingly, we can gain insight into the certificate of infeasibility. Geometrically, if a fictitious set of measurement results $\data{\vec{m}_{\mathrm{inc}}}$ is inconsistent, this means it \alert{lies outside of the image of $\mathcal{S}$}. A remarkable fact from convex geometry now comes into play: \alert{it is always possible to find a hyperplane which separates a point from a convex set}. This is somewhat obvious in one, two and three dimensions, and in fact it holds in all dimensions. The certificate of infeasibility in \alert{precisely the specification of such a separating hyperplane}, as depicted in Fig.~\ref{f:cert infeas}.

As we will see in many instances throughout this book, dual SDPs can often be interpreted as providing \alert{certificates} or \alert{witnesses} of certain properties, and these can be understood geometrically as specifying separating hyperplanes. 

\begin{exercises}
	\item \label{exc:convex set} Show that the set $\mathcal{M}$ of measurement results -- the image of $\mathcal{S}$ from \eqref{e:statespace} under the mapping \eqref{e:S to R mapping} -- is a convex set. That is, show that if two sets of results $\vec{m_1}$ and $\vec{m_2}$ are both in $\mathcal{M}$, then so is any set of results of the form $\vec{m}' = p \vec{m_1} + (1-p)\vec{m_2}$ for $0 \leq p \leq 1$. 
\end{exercises}

\subsection{Property estimation}\label{s:qse pe}

It can sometimes be the case that we are less interested in estimating the full state $\rho$ that led to the results $\data{\vec{m}}$, and more interested in knowing some property of it. If this property can be calculated by a linear function $f(\var{\rho})$, such as energy or magnetization, then we can find the range of values which are possible, by finding both the minimum and maximum value of this property that any quantum state $\var{\rho}$ can have. To do this, we can modify the feasibility SDP \eqref{eq: State Estimation} and introducing the objective function $f(\var{\rho})$, which is either minimised or maximised. 

As an example, we can estimate the maximum expected value of an unperformed measurement with  observable $\data{\tilde{M}}$ by solving the following SDP
\begin{subequations}\label{eq: property Estimation}
	\begin{align}
		\text{maximise} &\quad \tr(\data{\tilde{M}}\var{\rho}) \\
		\text{subject to} &\quad \tr(\data{M_x} \var{\rho}) = \data{m_x}& &x = 1,\ldots,N,\\
		&\quad\tr(\var{\rho}) = 1,\\
		&\quad \var{\rho} \opgeq 0. 
	\end{align}
\end{subequations}
By solving both this, and the analogous minimisation problem, we find the range of possible outcomes for the measurement $\data{\tilde{M}}$, consistent with the set of measurement results observed. As previously, there are three regimes that can be encountered. If the measurements $\data{\vec{M}}$ are tomographically complete, the feasible set is a unique state $\mathcal{F} = \var{\rho^*}$, and there will be a unique expectation value $\tilde{m} = \tr(\data{\tilde{M}}\var{\rho^*})$. Conversely, if there is no solution (because the measurement results $\data{\vec{m}}$ are inconsistent), the optimal value is $\alpha^* = -\infty$, signalling the infeasibility of the problem. Finally, when the measurements are consistent but not tomographically complete, we would expect there to be a range of values, all consistent with the measurement results. 

In Sections \ref{s:qse td} and \ref{s:qse fid} we saw that it is possible to calculate the trace distance and fidelity to a target state $\data{\sigma}$. From the current perspective, we can view these are two further instances of property testing -- of the state having the property of being $\sigma$. As was seen previously, even though the trace distance and fidelity of non-linear functions, both of these properties can nevertheless be estimated using semidefinite programming. It is possible to estimate other properties -- specified by non-linear functions -- using similar techniques. 

\section{The quantum marginal problem}\label{s:marginal problem}
In this final section we will now consider the quantum marginal problem. This problem is closely related to quantum state estimation, with the key difference being in the type of information given. In the previous sections of this chapter, we considered being given a set of measurement results, and were interested in estimating the quantum state -- or properties of it -- from this measurement data. We discussed in those sections the idea of tomographically complete data, \ie data which is sufficient to uniquely determine the quantum state. 

In the context of the quantum marginal problem, we will be interested in \alert{multipartite} quantum states, comprising a number of subsystems. The assumption now made is that we have perfect knowledge of some of the subsystems -- \ie some of the \alert{marginal} states, more commonly referred to as reduced density operators. This perfect knowledge can be thought of as being obtained by having tomographically complete data for these subsystems -- \ie by having performed a tomographically complete set of measurements, and having the corresponding measurement results. The basic problems of the quantum marginal problem are then to either determine the \alert{global state} consistent with all of the marginals, or to estimate some property of this global state, given access only to the marginals. 

We could now phrase everything just as was done above, in terms of measurement results. However, since the focus is now exclusively on situations where we have full knowledge of a number of subsystems, it is useful to adopt a more direct approach, and simply specify the reduced density operators of those subsystems. On the one hand, this is a conceptually cleaner way to approach the problem. On the other hand, it is an important lesson to realise that even in this formulation, we can still apply the techniques of semidefinite programming. 

As a concrete example, let us consider the direct quantum analogue of the (classical) marginal problem considered in Example 1.11. Therefore, consider a tripartite quantum system, with subsystems labelled by $X$, $Y$ and $Z$. Suppose that we have knowledge of the bipartite reduced density operators $\data{\rho_{XY}}$, $\data{\rho_{XZ}}$ and $\data{\rho_{YZ}}$, but not of the joint state $\var{\rho_{XYZ}}$. Consider that we want to determine whether there is any joint state consistent with the reduced density operators, and to find an example of one if it does. This can be cast as the following feasibility SDP:
\begin{subequations}
	\begin{align}
		\text{find}&\quad \var{\sigma_{XYZ}}\\
		\text{subject to}&\quad \tr_Z(\var{\sigma_{XYZ}}) = \data{\rho_{XY}},\\
		&\quad \tr_Y(\var{\sigma_{XYZ}}) = \data{\rho_{XZ}},\\
		&\quad \tr_X(\var{\sigma_{XYZ}}) = \data{\rho_{YZ}},\\
		&\quad \var{\sigma_{XYZ}} \opgeq 0, \quad \tr(\var{\sigma_{XYZ}}) = 1. 
	\end{align}
\end{subequations}
As with the state estimation problem, there are a number of interesting variants of this basic problem that can be considered. First, we can consider both trace distance and fidelity estimation of the global state (or a marginal state) to a target state, in analogy to the problems studied in Sections~\ref{s:qse td} and \ref{s:qse fid}.  We study these generalisations in Exercises~\ref{exc:td marg} and \ref{exc:fid marg} respectively. Second, we can also consider the analogous problem of property estimation from Section~\ref{s:qse pe}. This is studied in Exercise~\ref{exc:marginal property} below. 

We can also consider a problem which is closely related to the finite-statistics version of state estimation from Section~\ref{s:qse fs}. Here we imagine that instead of knowing the bipartite reduced density operators exactly, they are only known approximately. One way to model this is to assume that the marginals need to be $\epsilon$-close to a fixed state, \ie to demand
\begin{equation}
	\| \var{\sigma_{XY}} - \data{\rho_{XY}} \|_1 \leq \data{\epsilon},
\end{equation}
where we have used the shorthand $\var{\sigma_{XY}} = \tr_Z(\var{\sigma_{XYZ}})$, and similarly for $\var{\sigma_{XZ}}$ and $\var{\sigma_{YZ}}$. We note that, just as when considering finite statistics, $\data{\epsilon}$ is considered as being data which is specified in the problem -- allowing us to impose either stricter or weaker constraints on how close the reduced density operators of $\var{\rho_{XYZ}}$ need to be to the target states. Thus, we arrive at the following optimisation problem
\begin{subequations}\label{e:QMP norm orig}
	\begin{align}
		\text{find}&\quad \var{\sigma_{XYZ}}\\
		\text{subject to}&\quad \| \var{\sigma_{XY}} - \data{\rho_{XY}} \|_1 \leq \data{\epsilon},\label{e:QMP fs1}\\
		&\quad \| \var{\sigma_{XZ}} - \data{\rho_{XZ}} \|_1 \leq \data{\epsilon},\\
		&\quad \| \var{\sigma_{YZ}} - \data{\rho_{YZ}} \|_1 \leq \data{\epsilon},\label{e:QMP fs3}\\
		&\quad \var{\sigma_{XYZ}} \opgeq 0, \quad \tr(\var{\sigma_{XYZ}}) = 1. 
	\end{align}
\end{subequations}
As written, this is not an SDP, due to the non-linear inequality constraints \eqref{e:QMP fs1} -- \eqref{e:QMP fs1}. However, we can make use of the SDP characterisation of the $\epsilon$ trace norm ball, as given in Exercise 2.17, itself a small extension of the unit trace norm ball from equation 2.36. In particular, we arrive at
\begin{subequations}
	\begin{align}
		\text{find}&\quad \var{\sigma_{XYZ}}\\
		\text{subject to}&\quad \var{\sigma_{XY}} - \data{\rho_{XY}} = \var{\omega_{XY}} - \var{\zeta_{XY}},&& \tr(\var{\omega_{XY}} + \var{\zeta_{XY}}) = \data{\epsilon},& &\var{\omega_{XY}} \opgeq 0,\; \var{\zeta_{XY}}\opgeq 0, \\
		&\quad \var{\sigma_{XZ}} - \data{\rho_{XZ}} = \var{\omega_{XZ}} - \var{\zeta_{XZ}},&&
		\tr(\var{\omega_{XZ}} + \var{\zeta_{XZ}}) = \data{\epsilon},& &\var{\omega_{XZ}} \opgeq 0,\; \var{\zeta_{XZ}}\opgeq 0, \\
		&\quad \var{\sigma_{YZ}} - \data{\rho_{YZ}} = \var{\omega_{YZ}} - \var{\zeta_{YZ}},&&
		\tr(\var{\omega_{YZ}} + \var{\zeta_{YZ}}) = \data{\epsilon},& &\var{\omega_{YZ}} \opgeq 0,\; \var{\zeta_{YZ}}\opgeq 0, \\
		&\quad \var{\sigma_{XYZ}} \opgeq 0, &&\tr(\var{\sigma_{XYZ}}) = 1. 
	\end{align}
\end{subequations}
In order to emphasise the relationship between this formulation and \eqref{e:QMP norm orig}, we have presented four constraints per line, which collectively replace the corresponding trace norm constraint from the former. Note also that we have had to introduce two new variables per constraint, thus arriving at a larger problem. Nevertheless, this form is explicitly an SDP, and shows that it is possible to solve the analogue of the finite statistics problem for the quantum marginal problem. Finally, although we chose to place a constraint in terms of the trace norm, it is also possible in principle impose any distance-type constraint, as long as it can be expressed as an SDP. Examples of such include other norms, such as the operator norm, as well as the fidelity, both of which are left as exercises below. 

Finally, we can also consider turning the problem from a feasibility SDP into a (standard) optimisation SDP. As always, there are numerous approaches that one can take to achieve this. Here we will consider a particularly simple approach which can be used when the target marginal states are \alert{pure}. In this case, a relaxation of the marginal problem is to find a global state whose marginals have the largest average fidelity with the target pure marginals. 

As a particular example, let us consider the analogue situation from Exercise 1.9 (c), where only two of the marginals are specified.  Let us therefore assume that $\data{\rho_{XY}} = \data{\ket{\psi_{XY}}\bra{\psi_{XY}}}$ and $\data{\rho_{YZ}} = \data{\ket{\psi_{YZ}}\bra{\psi_{YZ}}}$. The largest average fidelity that can be achieved with these two states is given by the following simple SDP
\begin{subequations}\label{e:QMP simple}
	\begin{align}
		\text{maximise}&\quad \frac{1}{2} \left(\data{\bra{\psi_{XY}}}\var{\sigma_{XY}}\data{\ket{\psi_{XY}}} + \data{\bra{\psi_{YZ}}}\var{\sigma_{YZ}}\data{\ket{\psi_{YZ}}}\right)\\
		\text{subject to}&\quad \var{\sigma_{XYZ}} \opgeq 0,\quad \tr(\var{\sigma_{XYZ}}) = 1.
	\end{align}
\end{subequations} 
In Exercise 1.9 (c) it was shown that for the \alert{classical} version of this problem, with only two pairwise marginals specified, as long as they were compatible -- agreeing on the marginal distribution of the common random variable -- then a global distribution always exists which perfectly reproduces the two marginals. In the quantum setting, interestingly, this is no longer the case. In particular,  given two states $\data{\ket{\psi_{XY}}}$ and $\data{\ket{\psi_{YZ}}}$ such that $\data{\rho_Y} = \tr_X (\data{\ket{\psi_{XY}}}\data{\bra{\psi_{XY}}}) = \tr_Z (\data{\ket{\psi_{YZ}}}\data{\bra{\psi_{YZ}}})$, in general there \alert{will not} be a joint state of the three particles. The origin of this is \alert{monogamy of entanglement} -- a particle cannot simultaneously be highly entangled with two other particles, which introduces completely novel constraints into the marginal problem. As will be seen below in Exercise~\ref{exc:QMP dual}, we can use the dual of \eqref{e:QMP simple} to show that a particle cannot simultaneously be in a pure entangled state with two other particles.

In Chapter 5 we will return to an important variant of the marginal problem when studying the notion of a \alert{k-symmetric extension}, a powerful tool in the theory of entanglement.
\begin{exercises}
	\item \label{exc:td marg} In this exercise we will consider problems involving finding the closest state to a target state -- either the global state, or a marginal -- in terms of the trace distance.\\
	(a) Consider a situation involving three particles, where we are told all of the pairwise marginals, $\data{\rho_{XY}}$, $\data{\rho_{XZ}}$ and $\data{\rho_{YZ}}$, and wish to estimate the trace distance between the closest compatible state $\var{\sigma_{XYZ}}$ and a target global state $\data{\rho_{XYZ}}$. Write down the optimisation problem analogous to \eqref{e:QSE td orig} that needs to be solved. \lesson{This problem will not be an SDP}. \\
	(b) Make use of (2.8) in order to re-express the optimisation problem from part (a) as an SDP. \\
	(c) Repeat the exercise from parts (a) and (b), assuming now instead that only the marginal states  $\data{\rho_{XY}}$ and $\data{\rho_{XZ}}$ are specified, and that the goal is to estimate the trace distance between the closest compatible marginal state $\var{\sigma_{YZ}}$ and a target marginal state $\data{\rho_{YZ}}$. 
	
	\item \label{exc:fid marg} In this exercise we will repeat the same calculations as in the previous exercise, except now, instead of optimising the trace distance, we now optimise the fidelity.\\
	(a) Consider a situation involving three particles, where we are told all of the pairwise marginals, $\data{\rho_{XY}}$, $\data{\rho_{XZ}}$ and $\data{\rho_{YZ}}$, and wish to estimate the fidelity between the closest compatible state $\var{\sigma_{XYZ}}$ and a target global state $\data{\rho_{XYZ}}$. Write down the optimisation problem that needs to be solved. \lesson{This problem will not be an SDP}. \\
	(b) Make use of \eqref{e:F2 gen} in order to re-express the optimisation problem from part (a) as an SDP. \\
	(c) Repeat the exercise from parts (a) and (b), assuming now instead that the only the marginal states  $\data{\rho_{XY}}$ and $\data{\rho_{XZ}}$ are specified, and that the goal is to estimate the fidelity between the closest compatible marginal state $\var{\sigma_{YZ}}$ and a target marginal state $\data{\rho_{YZ}}$. 
	
	\item \label{exc:marginal property} In this exercise we will consider \alert{property estimation} in the context of the marginal problem. Consider that we are told the Hamiltonian $\data{H}$ of three particles, and the three pairwise marginal states $\data{\rho_{XY}}$, $\data{\rho_{XZ}}$ and $\data{\rho_{YZ}}$. Our goal is to find the range of average energies that the system can have, consistent with these marginals. \\
	(a) Write down an SDP that evaluates the minimum possible average energy any global state $\var{\sigma_{XYZ}}$ can have, given the Hamiltonian $\data{H}$ and marginal states. \\
	(b) Write down an SDP that evaluates the maximum possible average energy any global state $\var{\sigma_{XYZ}}$ can have, given the Hamiltonian $\data{H}$ and marginal states. 
	
	\item \label{exc:QMP dual} In this exercise we will derive the dual SDP of \eqref{e:QMP simple} and use it to show that, when in a pure state, a qubit cannot simultaneously be  entangled with two other qubits.\\
	(a) Write down the Lagrangian associated to the primal SDP \eqref{e:QMP simple}, using $\dual{M_{XYZ}}$ and $\dual{\mu}$ as dual variables for the first and second constraint, respectively. \\
	(b) Identify the constraints that need to be satisfied by the dual variables in order that \\ \hphantom{xx} (i) The Lagrangian upper bounds the value of the primal objective function for all primal feasible variables.\\ \hphantom{xx} (ii) The Lagrangian becomes independent of the primal variables. \\
	(c) Use parts (a) and (b) to show that, after solving for the slack variables, the dual SDP to \eqref{e:QMP simple} is
	\begin{subequations}\label{e:QMP simple dual}
		\begin{align}
			\text{minimise} &\quad \dual{\mu} \\
			\text{subject to}&\quad \dual{\mu}\identity \opgeq \frac{1}{2}\left(\data{\ket{\psi_{XY}}\bra{\psi_{XY}}} \otimes \identity_Z + \identity_X \otimes \data{\ket{\psi_{YZ}}\bra{\psi_{YZ}}}\right) 
		\end{align}
	\end{subequations} 
	\lesson{This SDP can be solved explicitly, with optimal value given by $$ \tfrac{1}{2}\Big\|\data{\ket{\psi_{XY}}\bra{\psi_{XY}}} \otimes \identity_Z + \identity_X \otimes \data{\ket{\psi_{YZ}}\bra{\psi_{YZ}}}\Big\|_\infty.$$}
	(d) For two projection operators $\Pi_1$ and $\Pi_2$, the following bound can be shown to hold: \begin{equation}\label{e:proj bound} \| \Pi_1 + \Pi_2 \|_\infty \leq 1 + \sqrt{\|\Pi_2 \Pi_1 \Pi_2 \|_\infty }.\end{equation}
	Writing the states $\data{\ket{\Psi_{XY}}}$ and $\data{\ket{\Psi_{YZ}}}$ in terms of their Schmidt decompositions, 
	\begin{align*}
		\data{\ket{\Psi_{XY}}} &= \sum_i \data{\sqrt{p_i}\ket{i_X}\ket{i_Y}},& \data{\ket{\Psi_{YZ}}} &= \sum_i \data{\sqrt{q_i}\ket{i_Y}\ket{i_Z}},
	\end{align*}
	use \eqref{e:proj bound} to show that the optimum value of the dual SDP \eqref{e:QMP simple dual} is not larger than
	\begin{equation}
		\dual{\mu^*} \leq \frac{1}{2}\left(1 + \max_i \sqrt{\data{p_i q_i}} \right).
	\end{equation}
	(e) Explain why the answer to part (d) shows that a qubit cannot simultaneously be in a pure entangled state with two other particles. 
\end{exercises}

\section{Concluding remarks}

In this chapter we began our exploration of applying semidefinite programming techniques to concrete problems in quantum information science. We mainly focused on the problem of quantum state estimation, which allowed us to study a variety of interesting features of SDPs. In particular, the main results described in this chapter are:
\begin{itemize}
	\item \textbf{Quantum state estimation.} The first problem dealt with was determining the existence of a quantum state compatible with a set of measurement data. We introduced a few variants of this problem, in particular, focusing on different figures of merit, such as the trace distance and the fidelity. This taught us that problems involving non-linear objective functions can sometimes be cast as SDPs. 
	\item \textbf{Double minimisation.} We shows that it is possible to cast problems involving double minimisation or double maximisation as SDPs -- See \eqref{e:trace distance SDP data}.
	\item \textbf{Certificates of infeasibility.} When considering feasibility problems reformulated as optimisation problems, the dual variables of SDPs can often be viewed as providing certificates of infeasibility. This follows from the facts that (i) every feasibility SDP can be turned into an optimisation SDP in which a positive solution indicates that the original problem was infeasible; (ii) Due to weak duality, the value of the objective function of the dual SDP lower bounds the optimal value of the primal SDP. This allows us to analytically certify infeasibility. 
\end{itemize}

\section{Further reading}
\begin{itemize}
	\item \href{https://doi.org/10.1017/CBO9780511976667 }{\textit{Quantum Computation and Quantum Information}}, M. A. Nielsen and I. L. Chuang, Cambridge Univ. Press, 2000.
\end{itemize}

\section{Advanced topics}
\subsection{The fidelity SDP}\label{s:fidelity SDP}
In this section we will derive the SDP formulation for fidelity as given in \eqref{e:F2 gen}. Recall that for a general pair of quantum states $\data{\rho}$ and $\data{\sigma}$, the fidelity is defined by
\begin{equation}
	F(\data{\rho},\data{\sigma}) = \| \sqrt{\data{\rho}}\sqrt{\data{\sigma}}\|_1^2.
\end{equation}
Our starting point in this section will be \alert{Uhlmann's theorem}, which gives an alternative expression for the fidelity in terms of \alert{purifications} of $\data{\rho}$ and $\data{\sigma}$. Recall that a purification of a (mixed) state $\omega$ is a bipartite state $\ket{\chi}$ such that
\begin{equation}
	\tr_B[\ket{\chi}\bra{\chi}] = \omega,
\end{equation}
where we label the systems as $A$ and $B$. 

Uhlmann's theorem then gives a classification of the fidelity in terms of the biggest overlap between any purification of $\data{\rho}$ and $\data{\sigma}$, namely
\begin{subequations}
	\begin{align}
		F(\data{\rho},\data{\sigma}) = \text{maximise} &\quad |\var{\langle \psi | \phi \rangle}|^2\\
		\text{subject to} &\quad \tr_B[\var{\ket{\psi}\bra{\psi}}] = \data{\rho}, \\
		&\quad \tr_B[\var{\ket{\phi}\bra{\phi}}] = \data{\sigma}.
	\end{align}
\end{subequations}
Let us consider that we have found an arbitrary purification of $\data{\rho}$ and an arbitrary purification of $\data{\sigma}$, which we denote by $\ket{\psi}$ and $\ket{\phi}$ respectively. Consider then the following state
\begin{equation}\label{e:ent fid}
	\ket{\Psi} = \frac{1}{\sqrt{2}}\left(\ket{\psi}\ket{0} + e^{i\theta} \ket{\phi}\ket{1}\right),
\end{equation}
where we have now introduced a third system, labelled $C$, taken to be a qubit, and an arbitrary phase factor $e^{i\theta}$ which we will fix below.

The basic observation we make is that if $\ket{\psi}$ and $\ket{\phi}$ are \alert{similar} -- \ie if they are close to being the same state, or have a large overlap $|\langle \psi | \phi \rangle |$ -- then the state $\ket{\Psi}$ \alert{cannot be very entangled}. 

In particular, as will be seen later in Chapter 5, the entanglement of a pure state can be determined by the reduced density operator of one subsystem. A state is entangled if and only if the reduced density operator (on any subsystem) is mixed. In our case, the reduced density operator of $C$ is 
\begin{equation}
	\omega_C = \tr_{AB}[\ket{\Psi}\bra{\Psi}] = \frac{1}{2}\left(\ket{0}\bra{0} + \ket{1}\bra{1} + e^{-i\theta}\langle \phi | \psi \rangle \ket{0}\bra{1} +e^{i\theta} \langle \psi | \phi \rangle \ket{1}\bra{0}\right).
\end{equation}
Let us now choose the angle $\theta$ such that $e^{i\theta}\langle \psi | \phi\rangle$ is real and non-negative (and therefore equal to $e^{-i\theta}\langle \phi | \psi \rangle$ and, moreover $|\langle \psi | \phi \rangle |$). The purity of $\omega_C$ is then seen to be
\begin{equation}
	\tr[\omega_C^2] = \frac{1}{2}\left(1+ |\langle \psi | \phi \rangle |^2\right).
\end{equation}
This shows that when $\ket{\psi}$ and $\ket{\phi}$ are orthogonal, such that $|\langle \psi | \phi \rangle | = 0$, then the purity of $\omega_C$ is minimal, since $\omega_C = \frac{1}{2}\identity$ is the maximally mixed state in this case. On the other hand, when $|\langle \psi | \phi \rangle | = 1$ -- meaning that they are the same state -- then the purity of $\omega_C$ is maximal, and indeed $\omega_C = \ket{+}\bra{+}$ is a pure state in this case. What this shows is that, as claimed, the entanglement of $\ket{\Psi}$ depends upon the overlap of $\ket{\psi}$ and $\ket{\phi}$, and can therefore be used to \alert{measure} it. 

We will now see how to use the above insight in order to derive (and interpret) the SDP for fidelity. Let us look instead at the reduced density operator $\omega_{AC}$,
\begin{subequations}
	\begin{align}
		\omega_{AC} &=\tr_B[\ket{\Psi}\bra{\Psi}], \\
		&= \frac{1}{2}( \data{\rho} \otimes \ket{0}\bra{0} + \data{\sigma}\otimes \ket{1}\bra{1}  \nonumber \\
		&\hspace{5em}+ e^{-i\theta}\tr_B[\ket{\psi}\bra{\phi}] \otimes \ket{0}\bra{1}+ e^{i\theta}\tr_B[\ket{\phi}\bra{\psi}] \otimes \ket{1}\bra{0})
	\end{align}
\end{subequations}
where we used the fact that $\tr_B[\ket{\psi}\bra{\psi}] = \data{\rho}$ and $\tr_B[\ket{\phi}\bra{\phi}] = \data{\sigma}$. Being a density operator, we know that $\omega_{AC} \opgeq 0$. Moreover, it can be seen that trace of the Hermitian part of the off-diagonal block $e^{-i\theta}\tr_B[\ket{\psi}\bra{\phi}]$ is
\begin{equation}
	\tr\left[\frac{e^{-i\theta}\tr_B[\ket{\psi}\bra{\phi}] + e^{i\theta}\tr_B[\ket{\phi}\bra{\psi}]}{2}\right] = |\langle \psi | \phi \rangle |.
\end{equation}
Why is this relevant? Well, recall that, due to Uhlmann's theorem, $|\langle \psi | \phi \rangle |$, when maximised, is precisely $\sqrt{F(\data{\rho},\data{\sigma})}$. We have however just realised that $|\langle \psi | \phi \rangle |$ is also the trace of the Hermitian part of the off-diagonal block of $\omega_{AC}$. This provides us with a way of obtaining an \alert{upper bound} on the fidelity. The off-diagonal block $e^{-i\theta}\tr_B[\ket{\psi}\bra{\phi}]$ can be treated as a \alert{variable}, which we denote by $\var{Y} + i\var{Z}$ (with $\var{Y} = \var{Y}^\dagger$ and $\var{Z} = \var{Z}^\dagger$ Hermitian operators) and then consider maximising the trace of $\var{Y}$ (the Hermitian part), subject to the constraint that $\omega_{AC}$ is a valid density operator, namely
\begin{subequations}\label{e:fid SDP adv}
	\begin{align}
		\sqrt{F(\data{\rho},\data{\sigma})} \leq \text{maximise} &\quad \tr(\var{Y}) \\
		\text{subject to} &\quad \begin{pmatrix}\data{\rho} & \var{Y} + i \var{Z}\\ \var{Y} - i \var{Z} & \data{\sigma}\end{pmatrix} \opgeq 0.
	\end{align}
\end{subequations}
where we have written $\omega_{AC}$ as a block matrix, and have ignored the overall factor of $\frac{1}{2}$, which doesn't affect whether or not it is positive semidefinite. This is an upper bound, since we know that it is possible to take $\var{Y} + i\var{Z} = e^{-i\theta}\tr_B[\ket{\psi}\bra{\phi}]$, with $\ket{\psi}$ and $\ket{\phi}$ the optimal purifications of $\data{\rho}$ and $\data{\sigma}$ respectively (which achieve the fidelity through Uhlmann's theorem), and with $\theta$ such that $e^{i\theta}\langle \psi | \phi\rangle$ is real. However, since we have relaxed the problem, and are maximising, there is no guarantee that this is the optimal solution, and so might obtain a strictly larger number than the fidelity. 

In order to show that this isn't the case, we can now proceed in the other direction, to show that $\tr(\var{Y^*}) \leq \sqrt{F(\data{\rho},\data{\sigma})}$. Let us assume that \eqref{e:fid SDP adv} has been solved, and that optimal variables $\var{Y^*}$ and $\var{Z^*}$ have been found. This in particular means that we have found a density operator
\begin{equation}
	\omega_{AC}^* = \frac{1}{2}( \data{\rho} \otimes \ket{0}\bra{0} + \data{\sigma}\otimes \ket{1}\bra{1} + (\var{Y^*}+i\var{Z^*}) \otimes \ket{0}\bra{1}+ (\var{Y^*}-i\var{Z^*}) \otimes \ket{1}\bra{0}).
\end{equation}
Consider then a purification of $\omega_{AC}^*$, which we denote $\ket{\Psi^*}$ (with the purifying system labelled $B$). From this purification we can define two (normalised) states,
\begin{align}
	\ket{\psi^*} &= \sqrt{2}(\identity_{AB} \otimes \bra{0}_C)\ket{\Psi^*},& \ket{\phi^*} &= \sqrt{2}(\identity_{AB} \otimes \bra{1}_C)\ket{\Psi^*},
\end{align}
Crucially, these states can be seen to be purifications of $\data{\rho}$ and $\data{\sigma}$. Indeed, we see that
\begin{subequations}
	\begin{align}
		\tr_B[\ket{\psi^*}\bra{\psi^*}] &= 2 \tr_B[(\identity_{AB} \otimes \bra{0}_C)\ket{\Psi^*}\bra{\Psi^*}(\identity_{AB} \otimes \ket{0}_C)],\\
		&= 2 (\identity_{A} \otimes \bra{0}_C)\omega_{AC}^*(\identity_{A} \otimes \ket{0}_C)],\\
		&= \data{\rho},
	\end{align}
\end{subequations}
and similarly $\tr_B[\ket{\phi^*}\bra{\phi^*}] = \data{\sigma}$. (Note also that this confirms that $\ket{\psi^*}$ and $\ket{\phi^*}$ are normalised). Now, the overlap between $\ket{\psi^*}$ and $\ket{\phi^*}$ is
\begin{subequations}
	\begin{align}
		|\langle \psi^* | \phi^*\rangle| &= 2 \bra{\Psi^*} (\identity_{AB} \otimes \ket{0}\bra{1}_C)\ket{\Psi^*},\\
		&= 2|\tr[(\identity_A \otimes \ket{0}\bra{1}_C)\omega_{AC}^*]|,\\
		&= |\tr(\var{Y^*}) + i\tr(\var{Z^*})|,\\
		&= \sqrt{\tr(\var{Y^*})^2 + \tr(\var{Z^*})^2 }.
	\end{align}
\end{subequations}
So, from Ulhmann's theorem, since $\ket{\psi^*}$ and $\ket{\phi^*}$ are purifications of $\data{\rho}$ and $\data{\sigma}$, we know that the fidelity $F(\data{\rho}, \data{\sigma})$ must be \alert{at least} $|\langle \psi^*|\phi^*\rangle|^2$, and therefore we have the bound
\begin{equation}
	\sqrt{F(\data{\rho},\data{\sigma})} \geq \langle |\psi^*|\phi^*\rangle| = \sqrt{\tr(\var{Y^*})^2 + \tr(\var{Z^*})^2 }.
\end{equation}
However, comparing with \eqref{e:fid SDP adv}, we have $\tr(\var{Y^*}) \geq \sqrt{F(\data{\rho},\data{\sigma})}$. The only way that these two bounds can be satisfied is if $\sqrt{F(\data{\rho},\data{\sigma})} = \tr(\var{Y^*})$ and $\tr(\var{Z^*}) = 0$. 

That is, this shows that in \eqref{e:fid SDP adv} the inequality can be replied by an equality, and that in fact this SDP precisely evaluates to the fidelity between $\data{\rho}$ and $\data{\sigma}$. 

The derivation moreover is interesting since it sheds light on how to interpret the SDP. We can view the objective function as maximising the entanglement of a state of the form \eqref{e:ent fid}, which entangles purifications of $\data{\rho}$ and $\data{\sigma}$. Only if the states have a small fidelity (meaning they are close to being orthogonal) can there exist purifying states that then can be superposed and generate a large amount of entanglement. When the states have a large fidelity, meaning they are similar, it is not possible to find purifications of them which generate much entanglement by superposing them. 

This can be viewed as a type of \alert{trade-off} relation, between how similar states are, and how much their purifications can be entangled with an auxiliary system, and shows that the fidelity quantifies this trade-off.

\end{document}